\documentclass[twocolumn,aps,pra,eqsecnum]{revtex4-1}
\usepackage[latin9]{inputenc}
\usepackage{textcomp}
\usepackage{bm}
\usepackage{amstext}
\usepackage{esint}

\makeatletter
\@ifundefined{textcolor}{}
{%
 \definecolor{BLACK}{gray}{0}
 \definecolor{WHITE}{gray}{1}
 \definecolor{RED}{rgb}{1,0,0}
 \definecolor{GREEN}{rgb}{0,1,0}
 \definecolor{BLUE}{rgb}{0,0,1}
 \definecolor{CYAN}{cmyk}{1,0,0,0}
 \definecolor{MAGENTA}{cmyk}{0,1,0,0}
 \definecolor{YELLOW}{cmyk}{0,0,1,0}
 }

\usepackage{bm}

\makeatother

\begin{document}

\title{Probabilistic Q-function distributions in fermionic phase-space}

\author{Laura E. C. Rosales-Z\'arate and P. D. Drummond}

\affiliation{Centre for Quantum and Optical Science, Swinburne University of Technology,
Melbourne 3122, Australia}
\begin{abstract}
We obtain a positive probability distribution or Q-function for an
arbitrary fermionic many-body system. This is different to previous
Q-function proposals, which were either restricted to a subspace of
the overall Hilbert space, or used Grassmann methods that do not give
probabilities. The fermionic Q-function obtained here is constructed
using normally ordered Gaussian operators, which include both non-interacting
thermal density matrices and BCS states. We prove that the Q-function
exists for any density matrix, is real and positive, and has moments
that correspond to Fermi operator moments. It is defined on a finite
symmetric phase-space equivalent to the space of real, antisymmetric
matrices. This has the natural SO(2M) symmetry expected for Majorana
fermion operators. We show that there is a physical interpretation
of the Q-function: it is the relative probability for observing a
given Gaussian density matrix. The distribution has a uniform probability
across the space at infinite temperature, while for pure states it
has a maximum value on the phase-space boundary. The advantage of
probabilistic representations is that they can be used for computational
sampling without a sign problem. 
\end{abstract}
\maketitle

\section{Introduction}

Phase-space representations of the quantum mechanical density operator
were introduced by Wigner~\cite{Wigner_1932} and Moyal~\cite{Moyal_1949}.
Such mappings from quantum mechanics to phase-space have many physical
applications, especially in coherence theory~\cite{Glauber:1963}.
Positive phase-space distributions or Q-functions, introduced by Husimi~\cite{Husimi1940},
give mappings which allow quantum mechanical observables to be calculated
using probabilistic sampling.

This paper introduces a unique, positive fermionic phase-space representation.
This has the same useful properties as other Q-function methods which
have been used for bosonic and spin Hilbert spaces. The Q-function
derived here is a unique positive distribution, defined for all density
matrices, and applicable to fermionic many-body systems. Since it
is probabilistic, there are no sign problems when sampling the distribution. 

Examples of this general approach include the bosonic Q-function~\cite{Husimi1940},
the SU(2) Q-function~\cite{Arecchi:1972,Gilmore1975}, the positive
P-function~\cite{Drummond:1980}, and the general positive distributions
over Gaussian operators~\cite{Corney:2003,Corney:2004,Corney:2006_PRB}.
It is known that the sampling properties of Q-function methods scale
favorably in the limit of large systems, which allows high order correlation
functions to be computed for large spin systems~\cite{Reid2014Multi}.
Q-functions using atomic coherent states have also been used to study
the dynamics of superfluorescence~\cite{Haake:1979,Lee:1984,Drummond1984PhysLetts}.
Another application of this type of distribution, based on coherent
states for spin and oscillator states, has been used to study quantum
dynamics of the Dicke superfluorescence model~\cite{Altland:2012_NJP}
and thermalization processes~\cite{Altland_PRL2012_Qchaos}.

We introduce a general method for constructing such probability distributions,
which is not restricted to Fermi systems. Our definition is based
on the expectation value of a hermitian operator basis. In the present
case, this is the basis of Gaussian fermionic operators~\cite{Corney:2006_JPA,Corney:2006_PRB,Corney:2004}.
Importantly, we prove that these provide a resolution of unity~\cite{ResUnityFGO:2013},
and have simple differential identities for operator moments. This
is fundamental to constructing a useful probabilistic representation. 

A Q-function provides a useful method for visualizing and understanding
coherence and correlations. It also has potential applications as
a computational tool. Since the distribution is always probabilistic,
there are no intrinsic sign problems with sampling or Monte Carlo
methods. We will treat detailed applications elsewhere. 

To achieve this result, we require the representation to satisfy the
following three properties, all found with the Husimi function: 
\begin{enumerate}
\item \emph{It exists uniquely for }any\emph{ quantum density-matrix.} 
\item \emph{It is a positive probability distribution.}
\item \emph{Observables are moments of the distribution.} 
\end{enumerate}
Can we satisfy these conditions in the case of fermions? 

Here, we obtain a fermionic Q-function which satisfies the three requirements
given above, and show how this fits into a general picture which includes
the well-known Husimi function. Our method makes use of a complete
set of positive-definite normally ordered operators as a basis~\cite{Corney:2006_JPA,Corney:2006_PRB,ResUnityFGO:2013},
and extends the conceptual basis of Q-function methods. The phase-space
is a bounded domain of real antisymmetric matrices. We show how the
theory of matrix polar coordinates and Riemannian measures on symmetric
spaces provides a natural mathematical framework for this probability.
Physically, we make use of the Class D symmetries introduced by Altland
and Zirnbauer~\cite{Altland_Zirnbauer:1997} to treat normal-superconducting
interfaces. We focus on this case, as it is the most general, while
bearing in mind that similar results will follow in each of the four
symmetry classes of this type.

This paper is organized as follows. In the next section we introduce
our general approach of defining a Q-function. In section (\ref{sec:Fermionic-Gaussian})
we describe the general normally ordered Gaussian operators and their
properties. In section (\ref{sec:Resolution-of-unity}) a completeness
proof is given, based on matrix polar coordinate theory. These results
are used to define the Q-function in section (\ref{sec:Fermionic-Q-function-definition}),
together with identities for observables and moments. In section (\ref{sec:Gaussian-density-operators}),
we give explicit expressions of the formalism developed here for Gaussian
density operators. Section (\ref{sec:Summary}) gives a summary of
our results and conclusions. Finally, properties of the unitary transformations
that we use are proved in Appendix (\ref{sec:Invariance_PC}) and
the inner product of the general Gaussian operators is calculated
in Appendix (\ref{sec:Appendix-B:-Inner}).

\section{Generalized Q-function properties\label{sec:Generalised-Q-functions}}

We first introduce a general definition of a positive Q-function representation
for \emph{any} Hilbert space. The approach given here is not just
applicable to fermions. It is therefore useful to understand the abstract
concepts first. Our approach differs in an essential way from that
of Husimi~\cite{Husimi1940} and others~\cite{Arecchi:1972,Perelomov_book_Coherent_state}.
Rather than constructing the phase-space representation using coherent
states, we employ a hermitian operator basis. This operator basis
can be thought as a basis of coherent \emph{operators}, rather than
coherent \emph{states}. It is closely related to the more general
idea of a Stratonovich-Weyl mapping~\cite{Stratonovich1956,Weyl_book_groups}.

\subsection{Q-function definition}

Suppose we have a positive definite, hermitian operator basis $\hat{\Lambda}\left(\vec{\lambda}\right)$
defined in a Hilbert space $\mathcal{H}$ of quantum mechanical operators,
where $\vec{\lambda}$ is a vector in the phase-space domain $\mathcal{D}$.
We require the following completeness property: the identity operator
$\hat{I}$ of the Hilbert space can be resolved as an integral over
the phase-space, so that
\begin{equation}
\int_{\mathcal{D}}\hat{\Lambda}\left(\vec{\lambda}\right)d\mu\left(\vec{\lambda}\right)=\hat{I}\,.\label{eq:Identity}
\end{equation}
This is called a resolution of unity. Here $d\mu\left(\vec{\lambda}\right)$
is an associated integration measure on the phase-space. A generalized
Q-function is defined as the inner product of the density matrix $\hat{\rho}$
with the operator basis:
\begin{equation}
Q\left(\vec{\lambda}\right)={\rm Tr}\left[\hat{\Lambda}\left(\vec{\lambda}\right)\hat{\rho}\right].\label{eq:Qdef}
\end{equation}

With this definition, condition (1) is automatically satisfied. We
will show below that condition (2) is also satisfied. Hence, a Q-function
exists for any density matrix in a Hilbert space which has a continuous,
positive resolution of the identity operator, proved suitable identities
are found that satisfy condition (3). 

There is a direct physical interpretation. The operators $\hat{\Lambda}\left(\vec{\lambda}\right)$
are observables, and the distribution simply gives their measured
expectation values. In the next sections, we demonstrate how the conditions
given above can be satisfied in the case of fermions.

\subsection{Probability distribution}

To demonstrate that the generalized Q-function defined above has the
properties of a probability distribution, as required for condition
(2), we must prove that it is positive and normalized for any density
matrix.

\paragraph{Positivity:}

Since the density matrix $\hat{\rho}$ is hermitian and positive definite,
it must have a diagonal Schmidt decomposition in an orthogonal basis
$\left|n\right\rangle $, as: $\hat{\rho}=\sum_{n}\rho_{n}\left|n\right\rangle \left\langle n\right|$,
where $\rho_{n}\ge0$. Since $\hat{\Lambda}\left(\vec{\lambda}\right)$
is positive definite by assumption, $\left\langle n\right|\hat{\Lambda}\left(\lambda\right)\left|n\right\rangle \ge0$,
so that
\begin{equation}
Q\left(\vec{\lambda}\right)={\rm Tr}\left[\hat{\Lambda}\left(\vec{\lambda}\right)\hat{\rho}\right]=\sum_{n}\rho_{n}\left\langle n\right|\hat{\Lambda}\left(\vec{\lambda}\right)\left|n\right\rangle \ge0
\end{equation}
Thus, the resulting distribution function is positive-semidefinite,
as required.

\paragraph{Normalization:}

From the definition given in Eq. (\ref{eq:Identity}), it is clear
that the trace of \emph{any} operator can be expressed as an integral
over the phase-space, since:
\begin{equation}
{\rm Tr}\left[\hat{O}\right]=\int{\rm Tr}\left[\hat{O}\hat{\Lambda}\left(\vec{\lambda}\right)\right]d\mu\left(\vec{\lambda}\right)\,.
\end{equation}
Choosing $\hat{O}=\hat{\rho}$, it follows that the distribution is
normalized to unity, since from Eq (\ref{eq:Qdef}):
\begin{equation}
1={\rm Tr}\left[\hat{\rho}\right]=\int Q\left(\vec{\lambda}\right)d\mu\left(\vec{\lambda}\right)\,.
\end{equation}

\subsection{Observables and moments}

Next, we wish to satisfy condition (3) for the general Q-function
case. This requires the evaluation of observables in the form of ${\rm Tr}\left[\hat{\rho}\hat{O}_{n}\right]=\left\langle \hat{O}_{n}\right\rangle $,
and it leads to nontrivial requirements on the basis set. Since the
eigenvalue methods used for the Husimi Q-function are not always available,
we look for a more general approach. Proving that these requirements
are satisfied in the case of fermions will require an understanding
of the differential properties of the fermionic Gaussian operators. 

We suppose that there are a complete set of identities that allow
all operator moments of physical interest $\hat{O}_{n}$ to be mapped
into differential operators, so that:
\begin{eqnarray}
\hat{O}_{n}\hat{\Lambda}\left(\vec{\lambda}\right) & = & \mathcal{D}_{n}\left(\partial_{\vec{\lambda}},\vec{\lambda}\right)\hat{\Lambda}\left(\vec{\lambda}\right)\nonumber \\
\hat{\Lambda}\left(\vec{\lambda}\right)\hat{O}_{n}^{\dagger} & = & \mathcal{D}_{n}^{*}\left(\partial_{\vec{\lambda}},\vec{\lambda}\right)\hat{\Lambda}\left(\vec{\lambda}\right).\label{eq:GeneralIdentities}
\end{eqnarray}

Here the second equation follows from the first, together with the
assumption that the basis set is hermitian. We use the convention
that in $\mathcal{D}_{n}\left(\partial_{\vec{\lambda}},\vec{\lambda}\right)$,
all differential operators are ordered to the left of any functions
of $\vec{\lambda}$. It then follows using the resolution of the identity,
Eq (\ref{eq:Identity}), that any observable in the form of an operator
moment can be represented as:
\begin{eqnarray}
\left\langle \hat{O}_{n}\right\rangle  & = & {\rm Tr}\left[\hat{\rho}\hat{O}_{n}\int\hat{\Lambda}\left(\vec{\lambda}\right)d\mu\left(\vec{\lambda}\right)\right]\nonumber \\
 & = & \int\mathcal{D}_{n}\left(\partial_{\vec{\lambda}},\vec{\lambda}\right)Q\left(\vec{\lambda}\right)d\mu\left(\vec{\lambda}\right)
\end{eqnarray}

Finally, provided $\mathcal{D}_{n}\left(\partial_{\vec{\lambda}},\vec{\lambda}\right)$
can be transformed via integration of the differentials into $O_{n}\left(\vec{\lambda}\right)=\mathcal{D}_{n}\left(0,\vec{\lambda}\right)$,
with vanishing boundary terms, it follows that:
\begin{eqnarray}
\left\langle \hat{O}_{n}\right\rangle  & = & \int Q\left(\vec{\lambda}\right)O_{n}\left(\vec{\lambda}\right)d\mu\left(\vec{\lambda}\right)=\left\langle O_{n}\left(\vec{\lambda}\right)\right\rangle _{Q}\,
\end{eqnarray}
If the differential mappings exist, and the distribution $Q\left(\vec{\lambda}\right)$
allows partial integration, observables can be calculated as a probabilistic
distribution of moments over $\lambda$. It is important to choose
a phase-space mapping such that $O_{n}\left(\vec{\lambda}\right)$
is efficiently computable.

The set of observables $\hat{O}_{n}$ needs to include only the \emph{physically
relevant} moments. If the Hamiltonian has conservation laws, we are
usually not interested in their invariant dynamics. This allows one
to reduce both the Hilbert space and phase-space dimensionality. Dimension
reduction proves useful in the case of the ground state properties
of the fermionic Hubbard model, which was numerically solved using
fermionic P-function methods using both translational and number-conservation
symmetries~\cite{Imada:2007_GBMC}.

\subsection{Bosonic Q-function}

Before examining the fermionic case, we now show that this definition
includes the well-known bosonic Q-function~\cite{Husimi1940}. In
this case the relevant operator basis is a normalized set of bosonic
coherent state projectors $\left|\vec{\alpha}\right\rangle \left\langle \vec{\alpha}\right|$,
which are also used in the Glauber-Sudarshan P-representation~\cite{Glauber:1963_P-Rep,Sudarshan_1963_P-Rep}.
These are defined in terms of the normalized bosonic coherent states
$\left|\vec{\alpha}\right\rangle $, which are eigenstates of the
bosonic annihilation operators, $\vec{a}=\left(\hat{a}_{1},\ldots\hat{a}_{M}\right)$
with eigenvalues $\vec{\alpha}=\left(\alpha_{1},\ldots\alpha_{M}\right)$:
\begin{equation}
\hat{\Lambda}_{B}\left(\vec{\alpha}\right)=\frac{1}{\pi^{M}}\left|\vec{\alpha}\right\rangle \left\langle \vec{\alpha}\right|\,.
\end{equation}

These match our definition: they give a resolution of the identity,
since they have the property~\cite{Glauber:1963_P-Rep} that
\begin{equation}
\frac{1}{\pi^{M}}\int\left|\vec{\alpha}\right\rangle \left\langle \vec{\alpha}\right|d^{2M}\vec{\alpha}=\hat{I}\,.
\end{equation}
 The relevant phase-space of $\vec{\lambda}$ is the $M-$dimensional
complex space $\mathcal{C}^{M}$, and the measure is the standard
Euclidean volume measure of $d^{2M}\vec{\alpha}=\prod_{j}d\alpha_{j}^{(x)}d\alpha_{j}^{(y)}$
where $\vec{\alpha}=\vec{\alpha}^{(x)}+i\vec{\alpha}^{(y)}$. From
the definition in Eq (\ref{eq:Qdef}) above, one can express the Husimi
Q-function as a unique positive distribution:
\begin{equation}
Q\left(\vec{\alpha}\right)=\frac{1}{\pi^{M}}\left\langle \vec{\alpha}\right|\hat{\rho}\left|\vec{\alpha}\right\rangle .
\end{equation}

In this case, the mappings to observables are also straightforward.
For antinormally ordered operators of form $\hat{O}_{\mathbf{m}\mathbf{n}}=\prod\hat{a}_{i}^{m_{i}}\prod\hat{a}_{i}^{\dagger n_{i}}$,
the corresponding function on phase-space is a c-number moment $O_{\mathbf{m}\mathbf{n}}\left(\vec{\alpha}\right)=\prod\alpha_{i}^{m_{i}}\prod\alpha_{i}^{*n_{i}}$,
so that 
\begin{equation}
{\rm Tr}\left[\hat{O}_{\mathbf{m}\mathbf{n}}\hat{\rho}\right]=\int O_{\mathbf{m}\mathbf{n}}\left(\vec{\alpha}\right)Q\left(\vec{\alpha}\right)d^{2M}\vec{\alpha}\,.
\end{equation}
This result is obtained, as usual, from the application of the eigenvalue
equation for the annihilation operators and the definition of the
Q-function.

\section{Fermionic Gaussian operators\label{sec:Fermionic-Gaussian}}

There have been several previous approaches that have reproduced some,
but not all of the properties described in the Introduction. One approach~\cite{Bowden:1972}
was to directly introduce an $SU\left(2\right)$ based fermion coherent
state. This satisfied (1) and (2), but not (3). Subsequently, a Q-function
based on fermionic coherent state projectors was introduced using
$U(N)$ Lie group methods~\cite{Perelomov_1972_Coherent_states_LieG,Gilmore:1974,Gilmore1975,Perelomov_GCS_bosons,Klauder:1985}.
This satisfied (2), but not (1) and (3), since the projectors are
not a complete basis, and observables were not derived. A third approach
due to Cahill and Glauber~\cite{Cahill:1999}, used Grassmann coherent
states. This approach satisfies (1) and (3) but not (2), as a Q-function
defined this way is Grassmann valued, and therefore neither real nor
positive.

Here we define the fermionic Q-function using the method given above,
with fermionic Gaussian operators~\cite{Corney:2006_JPA}as a basis.
These are also utilized in the complementary fermionic P-function
representations~\cite{Corney:2006_PRB}, which are non-unique mappings
to a phase-space of larger dimension. These have been utilized for
evaluating the ground state of the fermionic Hubbard model, via Monte-Carlo
techniques~\cite{Imada:2007_GBMC}. Other applications of these fermionic
P-functions are to the quantum dynamics of Fermi systems, like molecular
dissociation~\cite{Ogren:2010_Qdynamics_fermions_MolDiss,Ogren2011_fermiondynamics},
and the linear entropy in a quantum phase space~\cite{PRA_Entropy_paper_2011}.

\subsection{Quadratic Hamiltonians}

Consider a fermionic system composed of $M$ modes. We define $\hat{\bm{a}}$
as a vector of $M$ annihilation operators and $\hat{\bm{a}}^{\dagger}$
as vector of $M$ creation operators, where $\hat{a}_{i}$ and $\hat{a}_{j}^{\dagger}$
obey the fermionic anticommutation relations: 
\begin{eqnarray}
\left\{ \hat{a}_{i},\hat{a}_{j}^{\dagger}\right\}  & = & \delta_{ij}\nonumber \\
\left\{ \hat{a}_{i},\hat{a}_{j}\right\}  & = & 0
\end{eqnarray}
An extended vector of all $2M$ operators is written as $\hat{\underline{a}}=\left(\hat{\bm{a}}^{T},\hat{\bm{a}}^{\dagger}\right)^{T}$,
while the corresponding adjoint vector is $\hat{\underline{a}}^{\dagger}=\left(\hat{\bm{a}}^{\dagger},\hat{\bm{a}}^{T}\right)=\left(\hat{a}_{1}^{\dagger},\ldots,\hat{a}_{M}^{\dagger},\hat{a}_{1},\ldots,\hat{a}_{M}\right)$.
We denote $2M\times2M$ matrices as $\underline{\underline{H}}$,
and $M\times M$ matrices as $\mathbf{h}$.

The most general quadratic form of the fermion fields, expanded in
mode operators, is just the well-known Bogoliubov-de Gennes Hamiltonian:
\begin{equation}
\hat{H}=\sum_{ij=1}^{M}\left[h_{ij}a_{i}^{\dagger}a_{j}+\frac{1}{2}\left(\Delta_{ij}a_{i}^{\dagger}a_{j}^{\dagger}+\Delta_{ij}^{\dagger}a_{i}a_{j}\right)\right].
\end{equation}
This can written in a compact form as $\hat{H}=\frac{1}{2}\hat{\underline{a}}^{\dagger}\underline{\underline{H}}\hat{\underline{a}}$,
where the $2M\times2M$ matrix $\underline{\underline{H}}$ is defined
as:

\begin{equation}
\underline{\underline{H}}=\left[\begin{array}{cc}
\mathbf{h} & \bm{\Delta}\\
-\bm{\Delta}^{*} & -\mathbf{h}^{T}
\end{array}\right].\label{eq:I_GO-1-1}
\end{equation}

In order for $\hat{H}$ to be hermitian, one has the fundamental requirement
that $\mathbf{h}=\mathbf{h}^{\dagger}$ and that $\bm{\Delta}=-\bm{\Delta}^{T}$.
These conditions can be written~\cite{Altland_Zirnbauer:1997} as
conditions on $\underline{\underline{H}}$,
\begin{equation}
\underline{\underline{H}}=\underline{\underline{H}}^{\dagger}=-\underline{\underline{\Sigma}}\underline{\underline{H}}^{T}\underline{\underline{\Sigma}}\,,\label{eq:class-D symmetry}
\end{equation}
where: 
\begin{equation}
\underline{\underline{\Sigma}}=\left[\begin{array}{cc}
\mathbf{0} & \mathbf{I}\\
\mathbf{I} & \mathbf{0}
\end{array}\right].\label{eq:I_GO-1}
\end{equation}

A Hamiltonian with $\Delta=0$ is number-conserving, and describes
a non-interacting Fermi gas. With $\Delta\neq0$, this becomes a phase-dependent
form appropriate for describing normal-superconductor interfaces,
where Cooper pairs can tunnel through a barrier, and is also useful
in mean-field treatments of superconductors.

\subsection{Gaussian operators\label{sub:General-Gaussian-operators-1}}

Gaussian operators arise in many physical contexts. In analogy with
Gaussian distributions in probability theory, these are defined to
be exponentials of quadratics in the field operators~\cite{Corney:2004}.
Here, we will introduce a unit trace, hermitian Gaussian operator
proportional to $\exp\left(-\beta\hat{H}\right)$. Although these
also play the role of free-field canonical density matrices, the states
that can be represented are completely general. 

Our motivation is similar to Glauber's~\cite{Glauber:1963_P-Rep}
use of bosonic coherent state projectors, which are also Gaussian
operators~\cite{Corney:2003}. Just as with bosons, fermion states
with phase-dependent terms do not physically occur in isolated systems.
However, they describe coherence properties in the simplest possible
way, and provide a complete basis which can be used to represent any
density matrix.

It is convenient to parametrize such Gaussian operators as unit-trace
normally ordered forms, defined as~\cite{Corney:2006_JPA,Corney:2006_PRB}:
\begin{equation}
\hat{\Lambda}\left(\underline{\underline{\sigma}}\right)=\sqrt{\det\left[i\underline{\underline{\sigma}}\right]}\hat{\Lambda}^{u}\left(\underline{\underline{\sigma}}^{-1}-2\underline{\underline{\bar{I}}}\right)\,,
\end{equation}
 where $\hat{\Lambda}^{u}$ is a normally-ordered but un-normalized
Gaussian operator,
\begin{eqnarray}
\hat{\Lambda}^{u}\left(\underline{\underline{\mu}}\right) & = & :\exp\left[-\hat{\underline{a}}^{\dagger}\underline{\underline{\mu}}\hat{\underline{a}}/2\right]:\,,\label{eq:Normalized_GGO}
\end{eqnarray}
and $\underline{\underline{\bar{I}}}$ is a diagonal matrix given
by:
\begin{equation}
\underline{\underline{\bar{I}}}=\left[\begin{array}{cc}
-\mathbf{I} & \mathbf{0}\\
\mathbf{0} & \mathbf{I}
\end{array}\right].\label{eq:I_GO}
\end{equation}
Here $\mathbf{0}$ and $\mathbf{I}$ are the $M\times M$ zero and
identity matrices, respectively. Normal ordering is denoted by $:\ldots:$,
hence $:\hat{a}_{i}\hat{a}_{j}^{\dagger}:=-\hat{a}_{j}^{\dagger}\hat{a}_{i}$,
while antinormal ordering is denoted by $\left\{ \ldots\right\} $,
hence $\left\{ \hat{a}_{j}^{\dagger}\hat{a}_{i}\right\} =-\hat{a}_{i}\hat{a}_{j}^{\dagger}$. 

The $2M\times2M$ matrix $\underline{\underline{\sigma}}=\left(\underline{\underline{\mu}}+2\underline{\underline{\bar{I}}}\right)^{-1}$
is the \emph{covariance} matrix, which has an identical symmetry to
$\underline{\underline{H}}$. In terms of an $M\times M$ hermitian
matrix $\mathbf{n}$ and a complex antisymmetric matrix $\mathbf{m}$,
one can write:
\begin{equation}
\underline{\underline{\sigma}}=\left[\begin{array}{cc}
\mathbf{n}^{T}-\mathbf{I} & \mathbf{m}\\
-\mathbf{m}^{*} & \mathbf{I}-\mathbf{n}
\end{array}\right].\label{eq:Sigma-decomposition}
\end{equation}
 In physical terms, the Gaussian operator $\hat{\Lambda}$ is simply
the normally ordered density operator of a finite temperature, noninteracting
Fermi gas with Hamiltonian $\underline{\underline{H}}$, where \textbf{n}
is the normal Green's function and $\mathbf{m}$ the anomalous Greens'
function. These density operators arise in the theory of non-interacting
Fermi gases and BCS superconductors. The relationship between the
$\underline{\underline{\sigma}}$ and $\underline{\underline{H}}$
matrices is known~\cite{Ma:1995}, as is the fact that the symmetry
properties of $\underline{\underline{\sigma}}$ and $\underline{\underline{H}}$
are the same. However, for our purposes, only the normally ordered
covariance matrix $\underline{\underline{\sigma}}$ is important,
as it will define the Q-function phase-space.

\subsection{Differential identities\label{sub:Differential-identities_GO}}

The normally ordered Gaussian operators have the advantage of having
simple differential identities, which will allow us to obtain the
Q-function observables. These correspond to the action of the extended
creation and annihilation operators on the Gaussian basis~\cite{Corney:2006_JPA,Corney:2006_PRB}:
\begin{eqnarray}
:\underline{\hat{a}}\,\hat{\underline{a}}^{\dagger}\hat{\Lambda}: & = & \underline{\underline{\sigma}}\hat{\Lambda}-\underline{\underline{\sigma}}\frac{\partial\hat{\Lambda}}{\partial\underline{\underline{\sigma}}}\underline{\underline{\sigma}}.\nonumber \\
\left\{ \hat{\underline{a}}:\hat{\underline{a}}^{\dagger}\hat{\Lambda}:\right\}  & = & -\underline{\underline{\sigma}}\hat{\Lambda}-\underline{\underline{\tilde{\sigma}}}\frac{\partial\hat{\Lambda}}{\partial\underline{\underline{\sigma}}}\underline{\underline{\sigma}}.\nonumber \\
\left\{ :\hat{\Lambda}\hat{\underline{a}}:\hat{\underline{a}}^{\dagger}\right\}  & = & -\underline{\underline{\sigma}}\hat{\Lambda}-\underline{\underline{\sigma}}\frac{\partial\hat{\Lambda}}{\partial\underline{\underline{\sigma}}}\underline{\underline{\tilde{\sigma}}}.\nonumber \\
\left\{ \hat{\underline{a}}\hat{\underline{a}}^{\dagger}\hat{\Lambda}\right\}  & = & -\underline{\underline{\tilde{\sigma}}}\hat{\Lambda}-\underline{\underline{\tilde{\sigma}}}\frac{\partial\hat{\Lambda}}{\partial\underline{\underline{\sigma}}}\underline{\underline{\tilde{\sigma}}}.\label{eq:Diff_Id}
\end{eqnarray}
Here $\tilde{\underline{\underline{\sigma}}}\equiv\underline{\bar{I}}-\underline{\underline{\sigma}}$.
We use nested ordering $\left\{ :\ldots:\right\} $ with the convention
that external orderings do not change orderings inside internal brackets,
and the ordering of $\hat{\Lambda}$is invariant. When calculating
matrix derivatives, we use the convention that: $\left[\partial/\partial\underline{\underline{\sigma}}\right]_{\alpha\beta}\equiv\partial/\partial\sigma_{\beta\alpha}$~\cite{Corney:2006_JPA}.
Due to matrix symmetry, one finds that
\begin{equation}
\frac{\partial\sigma_{\gamma\epsilon}}{\partial\sigma_{\alpha\beta}}=\delta_{\alpha\gamma}\delta_{\beta\epsilon}-\Sigma_{\alpha\epsilon}\Sigma_{\beta\gamma}.\label{eq:diff-matrix}
\end{equation}

\subsection{Group properties and transformations\label{sub:Group-properties-and}}

In order to prove our results in later sections, it is useful to relate
the Gaussian operators to fundamental results on the symmetry properties
of physical systems, and the concept of a symmetric space.

Dyson's `threefold way'~\cite{Dyson:1962_Threefold} classified the
symmetry groups of physical systems into real, complex and quaternion
types, corresponding to the Weyl classical groups~\cite{Weyl_book_groups}:
orthogonal, unitary and symplectic. The physical meaning of these
three classes relates to behavior under time reversal and spin rotation.
Additional, nonstandard symmetry groups have been found since, making
ten in all. The work of Altland and Zirnbauer~\cite{Altland_Zirnbauer:1997}
considered symmetry classes arising in the case of quadratic Hamiltonians
for coupled superconductor-normal fermionic systems. These are just
the transformations of the Gaussian fermionic operators defined above.

The most general case has neither time-reversal nor spin-rotation
invariance. This case, known as class D symmetry, allows one to treat
an arbitrary Fermi system. 

Many physical systems have a higher degree of symmetry than this.
This can be used to reduce the phase-space dimension, which has practical
advantages. For example, number conservation and translational symmetry
are utilized in Imada's analysis of the Hubbard model ground state~\cite{Imada:2007_GBMC},
which uses Gaussian operator methods. Such dimension reduction methods
can used for Q-functions also, but we do not treat them here for simplicity.

All Class D matrices have the symmetry indicated by Eq (\ref{eq:class-D symmetry}).
To obtain expressions which treat particles and holes on an equal
basis, we introduce the zeta matrices: 
\begin{equation}
\underline{\underline{\zeta}}=\underline{\underline{\bar{I}}}-2\underline{\underline{\sigma}}=\underline{\underline{\tilde{\sigma}}}-\underline{\underline{\sigma}}\,,
\end{equation}
which we will call `stretched' variables. These have an identical
class D symmetry. 

In order to understand their relationship with the classical symmetry
groups, we can transform the Fermi operators to an hermitian Majorana
fermion basis, $\hat{\underline{z}}=\underline{\underline{U_{0}}}\hat{\underline{a}}$.
This allows us to introduce a corresponding matrix mapping that preserves
quadratic forms:
\begin{equation}
\underline{\underline{X}}=i\underline{\underline{U_{0}}}\underline{\underline{\zeta}}\underline{\underline{U_{0}^{-1}}}\,,\label{eq:Antisymmetric-matrix transformation}
\end{equation}
where the unitary matrix $\underline{\underline{U_{0}}}$ is defined
as:
\begin{equation}
\underline{\underline{U_{0}}}=\frac{1}{\sqrt{2}}\left[\begin{array}{cc}
\mathbf{I} & \mathbf{I}\\
i\mathbf{I} & -i\mathbf{I}
\end{array}\right].\label{eq:U0}
\end{equation}

The Class D symmetry properties then become $\underline{\underline{X}}=\underline{\underline{X}}^{*}=-\underline{\underline{X}}^{T}$,
so these are $2M\times2M$ real antisymmetric matrices. Like the matrices
$i\underline{\underline{\zeta}}$, they form an $so(2M)$ Lie algebra.
This implies that the Gaussian operators $\hat{\Lambda}$ are a representation
of the $SO(2M)$ Lie group~\cite{Wybourne_Group_theory_book}. While
we focus mainly on the usual Fermi operators here, clearly there is
an equivalent approach using hermitian Majorana operators and the
real antisymmetric $\underline{\underline{X}}$ matrices.

An important consequence of this group theoretic correspondence~\cite{Altland_Zirnbauer:1997}
is that there is a unitary matrix $\underline{\underline{U}}$ that
diagonalizes any hermitian matrix $\underline{\underline{\zeta}}$,
while retaining the symmetry of Eq (\ref{eq:class-D symmetry}). This
unitary matrix is also a member of the $SO(2M)$ Lie group, with the
property that 
\begin{equation}
\underline{\underline{U}}^{-1\dagger}=\underline{\underline{U}}=\underline{\underline{\Sigma}}\underline{\underline{U}}^{*}\underline{\underline{\Sigma}}.\label{eq:diagonalize-unitary}
\end{equation}
This requirement means that the unitary transformation $\hat{\underline{b}}=\underline{\underline{U}}\hat{\underline{a}}$
preserves the Fermi commutators~\cite{Balian:1969}.

The preservation of Fermi commutators on the extended vector of operators
is essential to the unitary transformations in this paper. It is used
in Appendix~\ref{sec:Invariance_PC} to show that a normally-ordered
Gaussian operator remains normally-ordered in the new variables, after
a unitary transformation of both the operators and the covariance
matrix.

\subsection{Positivity}

\emph{While $\hat{\Lambda}$ is hermitian, is it also positive-definite?} 

Because not all Gaussian operators $\hat{\Lambda}$ of the form given
in (\ref{eq:Normalized_GGO}) are positive definite, only a finite
domain in the phase-space of $\underline{\underline{\zeta}}$ variables
leads to a physical density matrix with positive eigenvalues. To find
a necessary and sufficient restriction, we use the fact that the corresponding
covariance $\underline{\underline{\sigma}}$ matrix is hermitian,
and so is diagonalizable using the unitary transformation of Eq (\ref{eq:diagonalize-unitary})
to new operators \textbf{$\hat{\underline{b}}=\underline{\underline{U}}\hat{\underline{a}}$},
leaving the commutation relations invariant. 

We show in the Appendix that this diagonalizing transformation can
be applied inside the normally ordered symbols, and leaves invariant
the Class D reflection symmetry on the diagonals. After transforming
$\underline{\underline{\zeta}}$, we obtain
\begin{equation}
\underline{\underline{U}}\underline{\underline{\zeta}}\underline{\underline{U}}^{-1}=\left[\begin{array}{cc}
\mathbf{I}-2\mathbf{n}'\\
 & 2\mathbf{n}'-\mathbf{I}
\end{array}\right],
\end{equation}
where $\mathbf{n}'$ is diagonal. On re-ordering the anti-normal terms
in the exponential, which requires a sign-change owing to the definition
of fermionic normal ordering, the un-normalized Gaussian becomes:
\begin{equation}
\hat{\Lambda}^{u}=\prod_{j=1}^{M}\left[1+\hat{b}_{j}^{\dagger}\frac{2n'_{j}-1}{1-n'_{j}}\hat{b}_{j}\right]\,.
\end{equation}

Noting that the possible eigenvalues of $\hat{b}_{j}^{\dagger}\hat{b}_{j}$
are $0,1$, we see that the condition $\left(\mathbf{1}-\mathbf{n}'\right)\mathbf{n}'>0$
is both necessary and sufficient for positivity of $\hat{\Lambda}^{u}$
and hence $\hat{\Lambda}$. Since a unitary transformation does not
change the positivity of a matrix, this can be re-written as an equivalent
condition on $\underline{\underline{\zeta}}$, defining the phase-space
domain $\mathcal{D}$ as:
\begin{equation}
\underline{\underline{I}}-\underline{\underline{\zeta}}^{2}>0.\label{eq:stretched-variable-domain}
\end{equation}

The stretched $\underline{\underline{\zeta}}$ matrices therefore
have an eigenvalue range from $(-1,1)$. This domain corresponds to
the positive definite hermitian operators of interest.

\subsection{Classical domains and symmetric spaces}

Since the class D symmetric covariance matrices can be transformed
into real antisymmetric $2M\times2M$ matrices using Eq (\ref{eq:Antisymmetric-matrix transformation}),
the hermitian Gaussian operators can also be regarded as having a
real phase space of $M(2M-1)$ dimensions. The boundary of this space,
equivalent to the domain $\mathcal{D}$ of Eq (\ref{eq:stretched-variable-domain}),
is defined by the requirement:
\begin{equation}
\underline{\underline{I}}+\underline{\underline{X}}^{2}>0.\label{eq:classical real antisymmetric domain}
\end{equation}

This is the real subspace of the irreducible homogeneous bounded symmetric
domain $\mathcal{R}_{III}$ of complex skew-symmetric matrices~\cite{Cartan:1935},
called a classical domain in the theory of matrix polar coordinates~\cite{Hua:1963}.
More generally, all spaces of this type, with Riemannian measure,
have a one-to-one relationship with the ten simple Lie groups that
describe physical symmetries. These are called the symmetric spaces
\cite{Helgason:2001}. 

Geometrically, we note the following property in the present case:
\begin{equation}
\sum_{j}X_{ij}^{2}<1
\end{equation}

Thus, every individual element of $\underline{\underline{X}}$ is
bounded, since $\left|X_{ij}\right|<1$, and every row and column
is bounded by a corresponding hyperspherical shell. This shows that
the Class D group symmetry properties have a natural correspondence
to a phase-space with a finite boundary. We note here that there are
four similar types of symmetric space of this general class, with
differing symmetry properties \cite{Altland_Zirnbauer:1997}. In principle,
any of these can be used to coonstruct a different type of fermionic
Q-function that is appropriate to the relevant symmetry group.

\section{Resolution of unity\label{sec:Resolution-of-unity}}

In order to obtain the Q-function representation, the next step is
to obtain a resolution of unity for the normally-ordered fermionic
Gaussian operators expressed in terms of the $\underline{\underline{\zeta}}$
matrices. We follow a similar procedure to the parity argument used
previously to obtain the resolution of unity for the fermionic Gaussian
operators~\cite{ResUnityFGO:2013}. The main difference is that the
present identity requires an integral over a finite domain, due to
our choice of a normally ordered basis set.

\subsection{Riemannian volume}

To carry out integration on the phase-space, it is necessary to obtain
a volume measure. Here we follow the original approach of Hurwitz~\cite{Hurwitz:1897,Forrester}
and Hua~\cite{Hua:1963}, by using measures defined relative to an
invariant metric
\begin{equation}
ds^{2}={\rm Tr}\left(d\underline{\underline{X}}^{T}d\underline{\underline{X}}\right)={\rm Tr}\left(d\underline{\underline{\zeta}}^{\dagger}d\underline{\underline{\zeta}}\right)\,.
\end{equation}
 This gives a unified invariant measure over both the real and complex
matrices described above. Given a distance metric, $ds^{2}=g_{ij}dx_{i}dx_{j}$,
the corresponding Riemannian volume measure is $d\mu=\sqrt{\left|g\right|}\prod dx_{i}$~\cite{FyodorovMetric}. 

To start with, we define:
\begin{eqnarray}
dX & = & \prod_{1\text{\ensuremath{\le}}j<k\text{\ensuremath{\le}}2M}dX_{jk}\,,\\
d\zeta & = & \prod_{j=1}^{M}d\zeta_{jj}\prod_{1\leq j<k\leq M}d^{2}\zeta_{jk}d^{2}\zeta_{jk+M}\,\,,\nonumber 
\end{eqnarray}
 which are the Euclidean measures for the independent elements of
the real anti-symmetric matrix $\underline{\underline{X}}$, and the
class D complex matrices respectively. 

The volume elements can now be computed. In the antisymmetric case,
the metric and corresponding Riemannian measure are
\begin{eqnarray}
ds^{2} & = & 2\sum_{1\text{\ensuremath{\le}}j<k\text{\ensuremath{\le}}2M}dX_{jk}^{2}.\nonumber \\
d\mu\left(X\right) & = & 2^{M(M\text{\textminus}1/2)}dX\,.
\end{eqnarray}
For Class D hermitian matrices the metric and Riemannian measure are
larger, since:
\begin{eqnarray}
ds^{2} & = & 2\sum_{j=1}^{M}d^{2}\zeta_{jj}+4\sum_{1\leq j<k\leq M}\left[\left|d\zeta_{jk}\right|^{2}+\left|d\zeta_{jk+M}\right|^{2}\right]\,\nonumber \\
d\mu\left(\zeta\right) & = & 2^{2M(M\text{\textminus}3/4)}d\zeta\,.\label{eq:Riemannian-measure}
\end{eqnarray}
 Thus, we have now defined the phase-space and a volume measure for
the Q-function operator basis.

\subsection{Matrix polar coordinates}

Integration over the phase-space of matrix variables is simplified
by using matrix polar coordinates, which are commonly used in random
matrix theory~\cite{Hua:1963,Altland_Zirnbauer:1997,Forrester:2010}.
Since $i\underline{\underline{\zeta}}$ belongs to a Lie algebra,
and $\underline{\underline{\zeta}}$ is hermitian, it can be diagonalized,
see Section (\ref{sub:Group-properties-and}):
\begin{equation}
\underline{\underline{\zeta}}=\underline{\underline{U}}^{-1}\underline{\underline{\zeta}}^{D}\underline{\underline{U}}.\label{eq:Y_sigmabar-1}
\end{equation}
Here $\underline{\underline{U}}$ is an element of the $SO\left(2M\right)$
group and $\underline{\underline{\zeta}^{D}}={\rm diag}\left(\bm{\zeta},-\bm{\zeta}\right)$.
The eigenvalues must be in the range $-1<\zeta_{j}<1$, since the
domain is such that $\underline{\underline{1}}-\underline{\underline{\zeta}}^{2}>0$.
The Jacobian for the transformation from the coordinate $\underline{\underline{\zeta}}$
to polar coordinates $\left(\mathbf{\bm{\zeta}},\underline{\underline{U}}\right)$,
is given by~\cite{Altland_Zirnbauer:1997,Forrester:2010}:
\begin{equation}
d\mu\left(\zeta\right)=\underline{\underline{U}}^{\dagger}d\underline{\underline{U}}\Delta^{2}\left(\bm{\zeta}^{2}\right)d\mathbf{\bm{\zeta}}\,.\label{eq:Jacobian_ShiftedSigma-2}
\end{equation}

Here we have used the matrix polar coordinate~\cite{Altland_Zirnbauer:1997}
measure for class D symmetry, where $d\mathbf{\bm{\zeta}}=\prod_{j=1}^{M}d\zeta_{j}$
and $\Delta\left(\bm{\zeta}^{2}\right)$ is the Vandermonde determinant
defined as:
\begin{equation}
\Delta(\mathbf{\mathbf{\bm{\zeta}}}^{2})=\prod_{1\le i<j\leq M}\left(\zeta_{i}^{2}-\zeta_{j}^{2}\right).
\end{equation}

To evaluate the Riemannian unitary volume $C_{R}=\int\left(\mathbf{U}^{\dagger}d\mathbf{U}\right)$,
we use our previous technique~\cite{ResUnityFGO:2013} of evaluating
a Gaussian integral in both rectangular and polar coordinates over
an infinite domain. The integral is:
\begin{equation}
G_{M}=\int_{-\infty}^{\infty}d\mu\left(\zeta\right)\exp\left[-{\rm Tr}\left(\underline{\underline{\zeta}}\underline{\underline{\zeta}}^{\dagger}\right)/4\right]\,.\label{eq:Gaussian-integral}
\end{equation}

From Riemannian volume invariance, this Gaussian integral can be evaluated
by unitary transformation to an equivalent real antisymmetric matrix,
which gives a one-dimensional real Gaussian integral in each coordinate:
\begin{eqnarray}
G_{M} & = & \int_{-\infty}^{\infty}d\mu\left(X\right)\exp\left[-{\rm Tr}\left(\underline{\underline{X}}\underline{\underline{X}}^{T}\right)/4\right]\nonumber \\
 & = & 2^{M(M\text{\textminus}1/2)}\int_{-\infty}^{\infty}\exp\left[-\sum_{_{i>j}}^{2M}X_{ij}^{2}/2\right]dX\nonumber \\
 & = & \left(4\pi\right)^{M(M-1/2)}.\label{eq:Cartesian-gaussian-integral}
\end{eqnarray}

Evaluating this using matrix polar coordinates: 
\begin{eqnarray}
G_{M} & = & \int\underline{\underline{U}}^{\dagger}d\underline{\underline{U}}\int_{-\infty}^{\infty}\Delta^{2}\left(\bm{\zeta}^{2}\right)d\mathbf{\bm{\zeta}}\exp\left[-{\rm Tr}\left(\underline{\underline{Y}}\underline{\underline{Y}}^{\dagger}\right)/4\right]\,\nonumber \\
 & = & C_{R}\int_{-\infty}^{\infty}\Delta^{2}\left(\bm{\zeta}^{2}\right)d\mathbf{\bm{\zeta}}\exp\left[-\sum_{i=1}^{M}\zeta_{i}^{2}/2\right].
\end{eqnarray}
This has the form of a Mehta integral used in random matrix theory~\cite{Mehta:2004},
 and one obtains:
\begin{eqnarray}
G_{M} & = & C_{R}(2)^{M(M-1/2)}\prod_{j=1}^{M}j!\Gamma\left(j-1/2\right).\label{eq:GaussianIntPC}
\end{eqnarray}

Hence, on comparing with Eq (\ref{eq:Cartesian-gaussian-integral})
we find that the Riemannian unitary volume is given by: 
\begin{eqnarray}
C_{R} & = & \left(2\pi\right)^{M\left(M-\frac{1}{2}\right)}\prod_{j=1}^{M}\frac{1}{j!\Gamma\left(j-1/2\right)}.\label{eq:CR}
\end{eqnarray}

Since the matrices $\underline{\underline{H}}$ used in Ref.~\cite{ResUnityFGO:2013}
also have the class-D symmetry, we can relate this volume to the Euclidean
volume $C^{\mathbf{U}}$ for the class-D matrices derived previously~\cite{ResUnityFGO:2013}.
As expected from Eq (\ref{eq:Riemannian-measure}), the relation between
these two factors is: 
\begin{equation}
C_{R}=2^{M\left(2M-\frac{3}{2}\right)}C^{\mathbf{U}}.
\end{equation}

We can now evaluate the invariant volume  ${\cal V}=\int_{{\cal D}}d\mu\left(\zeta\right)$
of the classical domain using matrix polar coordinates. In this case
we have to evaluate an integral of the form:
\begin{eqnarray}
{\cal V} & = & C_{R}\int_{-1}^{1}\Delta^{2}\left(\bm{\zeta}^{2}\right)d\mathbf{\bm{\zeta}}.
\end{eqnarray}
In order to perform this integral we consider the following change
of variables: $x_{j}=\zeta_{j}^{2}$, hence $d\zeta_{j}=\frac{1}{2}x_{j}^{-1/2}dx_{j}$.
This allows the integral to have the form of a Selberg integral~\cite{Mehta:2004}
\begin{eqnarray}
\int_{-1}^{1}\Delta^{2}\left(\bm{\zeta}^{2}\right)d\mathbf{\bm{\zeta}} & = & \int_{0}^{1}\ldots\int_{0}^{1}\left|\Delta\left(\mathbf{x}\right)\right|^{2}\prod_{j=1}^{M}x_{j}^{-1/2}dx_{j}\nonumber \\
 & = & \prod_{j=1}^{M}\frac{j!\Gamma\left(j-\frac{1}{2}\right)\Gamma\left(j\right)}{\Gamma\left(M+j-\frac{1}{2}\right)}.
\end{eqnarray}

Using the result of $C_{R}$ given in Eq. (\ref{eq:CR}) we obtain
that the invariant volume ${\cal V}$ of the classical domain is:
\begin{eqnarray}
\mathcal{V} & = & \left(2\pi\right)^{M\left(M-1/2\right)}\prod_{j=1}^{M}\frac{\left(j-1\right)!}{\Gamma\left(M+j-\frac{1}{2}\right)}.\label{eq:Volume}
\end{eqnarray}

\subsection{Normalized basis}

We now wish to prove that a resolution of unity for a suitably normalized
Gaussian basis $\hat{\Lambda}^{N}\left(\underline{\underline{\zeta}}\right)$
is given by:
\begin{equation}
\hat{I}=\int_{\mathcal{D}}d\mu\left(\zeta\right)\hat{\Lambda}^{N}\left(\underline{\underline{\zeta}}\right).\label{eq:ResolutionUnity-sigma}
\end{equation}

The Q-function basis $\hat{\Lambda}^{N}\left(\underline{\underline{\zeta}}\right)$
we shall use is a function of the stretched phase-space coordinates
$\underline{\underline{\zeta}}$ on the classical domain of volume
$\mathcal{V}$, which vanishes at the domain boundaries. It is defined
as:
\begin{equation}
\hat{\Lambda}^{N}\left(\underline{\underline{\zeta}}\right)=\frac{1}{\mathcal{N}}\hat{\Lambda}\left(\frac{1}{2}\left[\underline{\underline{\bar{I}}}-\underline{\underline{\zeta}}\right]\right)S\left(\underline{\underline{\zeta}}^{2}\right).\label{eq:GaussianBasis}
\end{equation}
Here $S\left(\underline{\underline{\zeta}}^{2}\right)$ is an even,
positive scaling function, and $\mathcal{N}$ is a positive constant
introduced for normalization purposes. This basis is positive definite,
since the Gaussian operators $\hat{\Lambda}\left(\underline{\underline{\sigma}}\right)$
are positive definite~\cite{Corney:2006_JPA,Corney:2006_PRB}, inside
the classical domain. The function $S$ must be invariant under unitary
transformations, with a typical general form being:
\begin{eqnarray}
S\left(\underline{\underline{\zeta}^{2}}\right) & = & \det\left[\left(\underline{\underline{I}}-\underline{\underline{\zeta}}^{2}\right)^{k/2}\right]e^{-\left[s{\rm Tr}\underline{\underline{\zeta}}^{2}/4\right]}.\label{eq:Normfunction_P}
\end{eqnarray}

Using Grassmann variables, it is possible to prove that the normally
ordered Gaussian operators can be brought into diagonal form by the
unitary transformation of Eq (\ref{eq:Y_sigmabar-1}) that diagonalizes
the class-D hermitian matrix, and remain normally ordered in the new
operator basis. This is shown in Appendix (\ref{sec:Invariance_PC}). 

The trace normalization term for a Gaussian operator has an equal
number of positive and negative eigenvalues. As a result, in terms
of the eigenvalues of $\underline{\underline{\zeta}}$ and the transformed
operators $\hat{\underline{b}}=\underline{\underline{U}}\hat{\underline{a}}$,
the Gaussian operator is:
\begin{eqnarray}
\hat{\Lambda} & =\det\left[\frac{1}{2}\left(\mathbf{I}+\bm{\zeta}\right)\right] & :\exp\left[\hat{\underline{b}}^{\dagger}\left(\underline{\underline{\bar{I}}}-\left(\underline{\underline{\bar{I}}}-\underline{\underline{\zeta}}^{D}\right)^{-1}\right)\hat{\underline{b}}\right]:\,.\nonumber \\
\end{eqnarray}

In order to obtain the resolution of unity we therefore must prove
the following result, 
\begin{eqnarray}
\hat{I} & = & \frac{1}{\mathcal{N}}C_{R}\hat{I}_{\zeta}.\label{eq:Identity-in-radial-form}
\end{eqnarray}
Here we have used the matrix polar coordinate Jacobian, Eq (\ref{eq:Jacobian_ShiftedSigma-2}),
the unitary volume of Eq. (\ref{eq:CR}), and we introduce a radial
integral: 
\begin{equation}
\hat{I}_{\zeta}=\int_{\mathcal{\mathcal{V}}}\hat{\Lambda}\left(\bm{\zeta}\right)S\left(\bm{\zeta}^{2}\right)\Delta^{2}\left(\bm{\zeta}^{2}\right)d\bm{\zeta}\,.\label{eq:radial-integral}
\end{equation}

\subsection{Radial integral}

To complete the proof of the resolution of unity, and obtain the normalizing
factor $\mathcal{N}$, we now focus on the radial part. This corresponds
to the integral over the eigenvalues $\bm{\zeta}$. For simplicity
in evaluating integrals we will take $s=0$ in the normalizing function
$S\left(\bm{\zeta}^{2}\right)$. The normally ordered Gaussian operators
can be expressed as:
\begin{eqnarray}
\Lambda\left(\bm{\zeta}\right) & = & :\exp\left[2\hat{\bm{b}}^{\dagger}\left(\left(\mathbf{I}+\bm{\zeta}\right)^{-1}-\mathbf{I}\right)\hat{\bm{b}}\right]:\det\left[\frac{1}{2}\left(\mathbf{I}+\bm{\zeta}\right)\right]\nonumber \\
 & = & \prod_{j=1}^{M}\left[\frac{1}{2}\left(1+\zeta_{j}\right)-\hat{b}_{j}^{\dagger}\hat{b}_{j}\zeta_{j}\right].\label{eq:GO-eigenval-1}
\end{eqnarray}
Using the result of Eq (\ref{eq:GO-eigenval-1}) and Eq (\ref{eq:radial-integral}),
we can write the radial integral as:
\begin{eqnarray}
\hat{I}_{\zeta} & = & \int_{\mathcal{V}}\hat{\Lambda}\left(\bm{\zeta}\right)S\left(\bm{\zeta}^{2}\right)\Delta^{2}\left(\bm{\zeta}^{2}\right)d\bm{\zeta}\nonumber \\
 & = & 2^{-M}\int_{-1}^{1}d\bm{\zeta}\Delta^{2}\left(\bm{\zeta}^{2}\right)S\left(\bm{\zeta}^{2}\right)+\nonumber \\
 & + & \int_{-1}^{1}d\bm{\zeta}\Delta^{2}\left(\bm{\zeta}^{2}\right)S\left(\bm{\zeta}^{2}\right)\prod_{j=1}^{M}\zeta_{j}\left[\frac{1}{2}-\hat{b}_{j}^{\dagger}\hat{b}_{j}\right].\label{eq:RadIntS-1}
\end{eqnarray}
 The value of the integral $\hat{I}_{\zeta}$ will depend on the normalization
function. Expressed in terms of the eigenvalues $\bm{\zeta}$, this
is:
\begin{equation}
S\left(\bm{\zeta}^{2}\right)=\det\left[\left(\mathbf{I}-\bm{\zeta}^{2}\right)^{k}\right],
\end{equation}
which is an even function of the eigenvalues.

The second integral of Eq. (\ref{eq:RadIntS-1}) has terms proportional
to $\zeta_{j}$, which is an odd function of $\zeta_{j}$, while all
the other terms are an even function of $\zeta_{j}$. Hence from the
parity of these functions, the odd integrals vanish, so $\hat{I}_{\zeta}$
is proportional to a unit operator. 

The radial integral which must be evaluated is:
\begin{eqnarray}
I_{\zeta} & = & \int_{-1}^{1}d\bm{\zeta}\Delta^{2}\left(\bm{\zeta}^{2}\right)\prod_{j}\left(1-\zeta_{j}^{2}\right)^{k}.\label{eq:RadInt_NfPH-1}
\end{eqnarray}

Next we perform the following change of variables: $x_{j}=\zeta_{j}^{2}$,
hence $d\zeta_{j}=\frac{1}{2}x_{j}^{-1/2}dx_{j}$. This allows the
integral of Eq. (\ref{eq:RadInt_NfPH-1}) to be written in the form:
\[
I_{\zeta}=\int_{0}^{1}\ldots\int_{0}^{1}\left|\Delta\left(\mathbf{x}\right)\right|^{2}\prod_{j=1}^{M}x_{j}^{-1/2}\left(1-x_{j}\right)^{k}dx_{j},
\]
 which is another modified Selberg integral~\cite{Mehta:2004}. Therefore,
the radial integral of Eq. (\ref{eq:RadInt_NfPH-1}) is:
\begin{eqnarray}
I_{\zeta} & = & \prod_{j=1}^{M}\frac{j!\Gamma\left(j-\frac{1}{2}\right)\Gamma\left(k+j\right)}{\Gamma\left(k+M+j-\frac{1}{2}\right)}.\label{eq:IRad-1}
\end{eqnarray}
Using the value of $I_{\zeta}$ given in Eq. (\ref{eq:IRad-1}), together
with the unitary integral, we obtain a general resolution of unity
with a normalization constant $\mathcal{N}$ of the form:
\begin{eqnarray}
\mathcal{N} & = & \left(2\pi\right)^{M\left(M-1/2\right)}2^{-M}\prod_{j=1}^{M}\frac{\Gamma\left(k+j\right)}{\Gamma\left(k+M+j-\frac{1}{2}\right)}.\nonumber \\
\label{eq:normalization factor}
\end{eqnarray}

In the limit of $k=0$, this normalization constant is just the phase-space
volume $\mathcal{V}$ of the real classical domain, divided by the
number of many-body quantum states, $2^{M}$. This follows, since
the requirement for the resolution of unity, in the limit of $k,s\rightarrow0$,
is:
\begin{equation}
\mathcal{N}=2^{-M}C_{R}\int_{-1}^{1}d\bm{\zeta}\Delta^{2}\left(\bm{\zeta}^{2}\right)=2^{-M}\int_{\mathcal{V}}d\mu\left(\zeta\right)\label{eq:Rel_N_Vol}
\end{equation}
The integral on the right hand side is the volume $\mathcal{V}$ of
the real classical domain defined in Eq. (\ref{eq:Volume}). Hence
in the limit of a uniform normalization, we simply have 
\begin{equation}
\lim_{k,s\rightarrow0}{\cal N}=2^{-M}\mathcal{V}\,.
\end{equation}

A Monte-Carlo integration was carried out to verify this result, by
generating $10^{6}$ random antisymmetric matrices $\underline{\underline{X}}$
with $-1<X_{ij}<1$, and testing positivity from the eigenvalues.
This was in good agreement with Eqs. (\ref{eq:normalization factor})
and (\ref{eq:Volume}) for $k,\, M\leq3$. 

These results imply a simple physical interpretation of the resolution
of unity. We note that for an arbitary operator $\hat{O}$, one immediately
obtains from the $k\rightarrow0$ limit of the resolution, Eq (\ref{eq:ResolutionUnity-sigma}),
that: 
\begin{equation}
\frac{1}{2^{M}}Tr\left(\hat{O}\right)=\frac{1}{\mathcal{V}}\int_{\mathcal{D}}d\mu\left(\zeta\right)Tr\left(\hat{O}\hat{\Lambda}\left(\underline{\underline{\zeta}}\right)\right).\label{eq:average overlap}
\end{equation}

\emph{Hence,} \emph{the average overlap of any Fermi operator with
an orthonormal basis element equals its average overlap with a unit
trace Gaussian operator}.

\section{Fermionic Q-function and observables\label{sec:Fermionic-Q-function-definition}}

Following the method of Section~\ref{sec:Generalised-Q-functions},
the fermionic Q-function is now defined inside the volume $\mathcal{V}$
in terms of the normalized Gaussian basis, as:
\begin{equation}
Q\left(\underline{\underline{\zeta}}\right)={\rm Tr}\left[\hat{\rho}\hat{\Lambda}^{N}\left(\underline{\underline{\zeta}}\right)\right].\label{eq:Qf}
\end{equation}
 This is the Hilbert-Schmidt inner product of the density operator
$\hat{\rho}$ with the normalized Gaussian basis, $\hat{\Lambda}^{N}\left(\underline{\underline{\zeta}}\right)$,
defined in Eq. (\ref{eq:GaussianBasis}), and hence satisfies our
first requirement.

The second requirement was that the Q-function should be a probability
distribution which is normalized to unity. The density matrix $\hat{\rho}$,
is hermitian and positive definite. The general normally ordered Gaussian
operators defined in Eq. (\ref{eq:GaussianBasis}), are hermitian
since the matrix $\underline{\underline{\zeta}}$ is hermitian, and
they are also positive-definite~\cite{Corney:2006_JPA,Corney:2006_PRB}
within the classical domain. Thus, the fermionic Q-function is a non-negative
probability distribution, and it is normalized from the resolution
of unity.

As an example of a Q-function, the operator $\hat{I}/2^{M}$ is the
infinite temperature density matrix with unit trace. This has a constant
overlap with any unit trace Gaussian of $2^{-M}$, so that in the
limit of $k\rightarrow0$ one obtains $Q=1/\mathcal{V}$, as expected
for a uniform, normalized probability.

\subsection{Differential identities of the fermionic basis\label{sub:Differential-Id-FermionicBasis}}

The third required property for a Q-function is that observables are
moments of the distribution. In order to obtain this property, we
will use the known differential identities given in Eq (\ref{eq:Diff_Id}),
which correspond to the action of the extended creation and annihilation
operators on the Gaussian basis~\cite{Corney:2006_JPA,Corney:2006_PRB}.
In particular, we will make use of the identity,
\begin{equation}
\left\{ \hat{\underline{a}}:\hat{\underline{a}}^{\dagger}\hat{\Lambda}:\right\} =-\underline{\underline{\sigma}}\hat{\Lambda}-\underline{\underline{\tilde{\sigma}}}\frac{\partial\hat{\Lambda}}{\partial\underline{\underline{\sigma}}}\underline{\underline{\sigma}}.\label{eq:Moments_NO}
\end{equation}
Here the nested ordering $\left\{ :\ldots:\right\} $ is defined as:
\begin{eqnarray}
\left\{ \hat{a}_{\alpha}:\hat{a}_{\beta}^{\dagger}\hat{\Lambda}:\right\}  & = & \left(\begin{array}{cc}
\hat{a}_{i}\hat{a}_{j}^{\dagger}\hat{\Lambda} & \hat{a}_{i}\hat{\Lambda}\hat{a}_{j}\\
-\hat{a}_{j}^{\dagger}\hat{\Lambda}\hat{a}_{i}^{\dagger} & -\hat{\Lambda}\hat{a}_{j}\hat{a}_{i}^{\dagger}
\end{array}\right).
\end{eqnarray}

The derivative of the fermionic Gaussian basis $\hat{\Lambda}^{N}$
with respect to $\underline{\underline{\sigma}}$ is obtained by using
the product rule as well as this identity. Hence, the action of the
ladder operators on the Gaussian basis $\hat{\Lambda}^{N}$ for this
particular nested ordering is given by:
\begin{eqnarray}
\left\{ \hat{\underline{a}}:\hat{\underline{a}}^{\dagger}\hat{\Lambda}^{N}:\right\}  & = & -\underline{\underline{\sigma}}\hat{\Lambda}^{N}-\underline{\underline{\tilde{\sigma}}}\frac{\partial\hat{\Lambda}^{N}}{\partial\underline{\underline{\sigma}}}\underline{\underline{\sigma}}+\underline{\underline{\tilde{\sigma}}}\hat{\Lambda}^{N}\frac{\partial\ln S}{\partial\underline{\underline{\sigma}}}\underline{\underline{\sigma}}.\nonumber \\
\label{eq:DiffId_GON}
\end{eqnarray}

\subsection{Observables}

The extended creation and annihilation operators can be expressed
in normal or antinormal form. We will consider the antinormal form
of the observables, which are expressed as:
\begin{eqnarray}
\left\langle \left\{ \hat{\underline{a}}\hat{\underline{a}}^{\dagger}\right\} \right\rangle  & = & {\rm Tr}\left[\hat{\rho}\left(\begin{array}{cc}
\hat{\bm{a}}\hat{\bm{a}}^{\dagger} & \hat{\bm{a}}\hat{\bm{a}}^{T}\\
\hat{\bm{a}}^{\dagger T}\hat{\bm{a}}^{\dagger} & -\left(\hat{\bm{a}}^{\dagger T}\hat{\bm{a}}^{T}\right)^{T}
\end{array}\right)\right].\nonumber \\
\label{eq:DefObs}
\end{eqnarray}
Using the resolution of unity given in Eq. (\ref{eq:ResolutionUnity-sigma})
we can write the observables as:
\begin{eqnarray}
\left\langle \left\{ \hat{a}_{\alpha}\hat{a}_{\beta}^{\dagger}\right\} \right\rangle  & = & {\rm Tr}\left[\int d\underline{\underline{\zeta}}\hat{\rho}\left(\begin{array}{cc}
\hat{a}_{i}\hat{a}_{j}^{\dagger}\hat{\Lambda}^{N} & \hat{a}_{i}\hat{\Lambda}^{N}\hat{a}_{j}\\
\hat{a}_{i}^{\dagger}\hat{\Lambda}^{N}\hat{a}_{j}^{\dagger} & -\hat{\Lambda}^{N}\hat{a}_{j}\hat{a}_{i}^{\dagger}
\end{array}\right)\right].\nonumber \\
 & = & {\rm Tr}\left[\int d\underline{\underline{\zeta}}\hat{\rho}\left\{ \hat{a}_{\alpha}:\hat{a}_{\beta}^{\dagger}\hat{\Lambda}^{N}:\right\} \right].\label{eq:Obs_NO1}
\end{eqnarray}
Using the expression for the observables given in Eq. (\ref{eq:Obs_NO1})
as well as the differential identity of Eq. (\ref{eq:DiffId_GON})
and the definition of the Q-function given in Eq. (\ref{eq:Qf}),
we obtain the following expression for the observables:
\begin{eqnarray}
\left\langle \left\{ \hat{a}_{\alpha}\hat{a}_{\beta}^{\dagger}\right\} \right\rangle  & = & {\rm Tr}\left[\int d\underline{\underline{\zeta}}\left[-\sigma_{\alpha\beta}\hat{\Lambda}^{N}-\tilde{\sigma}_{\alpha\gamma}\frac{\partial\hat{\Lambda}^{N}}{\partial\sigma_{\kappa\gamma}}\sigma_{\kappa\beta}\right]\hat{\rho}\right]\nonumber \\
 & + & {\rm Tr}\left[\int d\underline{\underline{\zeta}}\left[\tilde{\sigma}_{\alpha\gamma}\hat{\Lambda}^{N}\frac{\partial\ln S}{\partial\sigma_{\kappa\gamma}}\sigma_{\kappa\beta}\right]\hat{\rho}\right]
\end{eqnarray}

Next, substituting the definition of the Q-function in this expression
gives:
\begin{eqnarray}
\left\langle \left\{ \hat{a}_{\alpha}\hat{a}_{\beta}^{\dagger}\right\} \right\rangle  & = & \int d\underline{\underline{\zeta}}\left[\tilde{\sigma}_{\alpha\gamma}\frac{\partial\ln S}{\partial\sigma_{\kappa\gamma}}\sigma_{\kappa\beta}-\sigma_{\alpha\beta}\right]Q\nonumber \\
 & - & {\rm Tr}\left[\int d\underline{\underline{\zeta}}\tilde{\sigma}_{\alpha\gamma}\frac{\partial\hat{\Lambda}^{N}}{\partial\sigma_{\kappa\gamma}}\sigma_{\kappa\beta}\hat{\rho}\right].
\end{eqnarray}
On integrating the third term by parts, the boundary term vanishes
due to the scaling factor $S$, for the corresponding bounded classical
domain. We now consider the explicit form of the normalization function
$S$ and take the limit of $k,s\rightarrow0$ to simplify the result.
The matrix derivatives are carried out on making use of Eq (\ref{eq:diff-matrix}),
which includes the class-D symmetry property.

Hence, after changing to the stretched variables, the normally ordered
Greens functions are given in a simple form by:
\begin{eqnarray}
\left\langle \left\{ \underline{\hat{a}}\underline{\hat{a}}^{\dagger}\right\} \right\rangle  & = & C_{M}\int_{\mathcal{\mathcal{V}}}\underline{\underline{\zeta}}Q\left(\underline{\underline{\zeta}}\right)d\underline{\underline{\zeta}}-\frac{1}{2}\underline{\underline{\bar{I}}}\,,\label{eq:Moments_NO-Der}
\end{eqnarray}
where $C_{M}=2M-1/2$ is a constant that depends on the dimensionality.
This process can be iterated to obtain higher-order correlations.

In the infinite temperature case of $Q=1/\mathcal{V}$, the odd integral
on the right-hand side vanishes, which means that $\left\langle \hat{a}_{i}^{\dagger}\hat{a}_{i}\right\rangle =\left\langle \hat{a}_{i}\hat{a}_{i}^{\dagger}\right\rangle =1/2$.
This is an expected result due to particle-hole symmetry.

\section{Gaussian density operators\label{sec:Gaussian-density-operators}}

In this section we give explicit expressions of fermionic Q-functions
in the case of Gaussian density operators. This includes the common
cases of a non-interacting thermal gas and a BCS state. More generally,
since the Gaussian operators are a complete basis for any density
matrix, this allows any Q-function to be calculated if the corresponding
fermionic P-function~\cite{Corney:2006_PRB} is known. For simplicity,
we will consider the limiting normalization with $k,s\rightarrow0$,
unless stated otherwise.

\subsection{Gaussian operator inner product}

Consider two normalized Gaussian operators, $\hat{\Lambda}(\underline{\underline{\sigma}})$
and $\hat{\Lambda}(\underline{\underline{\sigma}}')$. If we consider
$\hat{\Lambda}(\underline{\underline{\sigma}})$ as the Q-function
basis, and $\hat{\Lambda}(\underline{\underline{\sigma}}')$ as a
physical density matrix, then evaluating the Q-function reduces to
evaluating their standard Hilbert-Schmidt inner product:
\begin{equation}
F\left(\underline{\underline{\sigma}},\underline{\underline{\sigma}}'\right)={\rm Tr}\left(\hat{\Lambda}(\underline{\underline{\sigma}})\hat{\Lambda}(\underline{\underline{\sigma}}')\right).
\end{equation}

In Appendix B, this is carried out following methods previously used
to calculate linear entropy~\cite{PRA_Entropy_paper_2011}, together
with Grassmann coherent states. This gives the result that:
\begin{equation}
F=\det\left[\left(\bar{\underline{\underline{I}}}-\underline{\underline{\sigma}}'\right)\left(\bar{\underline{\underline{I}}}-\underline{\underline{\sigma}}\right)+\underline{\underline{\sigma}}'\underline{\underline{\sigma}}\right]^{1/2},
\end{equation}
which takes a simpler form in terms of the stretched variables:
\begin{equation}
F=2^{-M}\sqrt{\det\left[\underline{\underline{I}}+\underline{\underline{\zeta}}'\underline{\underline{\zeta}}\right]}.\label{eq:stretched variable overlap}
\end{equation}

If the anomalous terms vanish, so that $\mathbf{m}=\mathbf{m}'=0$,
then one obtains:
\begin{equation}
\underline{\underline{\zeta}}'\underline{\underline{\zeta}}=\left[\begin{array}{cc}
\left(\mathbf{I}-2\mathbf{n}'\right)^{T}\left(\mathbf{I}-2\mathbf{n}^{T}\right)\\
 & \left(2\mathbf{n}'-\mathbf{I}\right)\left(2\mathbf{n}-\mathbf{I}\right)
\end{array}\right],
\end{equation}
and this reduces to a previously obtained result~\cite{PRA_Entropy_paper_2011}
of:
\begin{equation}
F=\det\left[\mathbf{n}\mathbf{n}'+\left(\mathbf{I}-\mathbf{n}\right)\left(\mathbf{I}-\mathbf{n}'\right)\right].
\end{equation}

In the case of a general coordinate $\underline{\underline{\zeta}}$,
it is instructive to consider the overlap of a density matrix with
itself. If we are on the limits of the classical domain, then $\underline{\underline{I}}-\underline{\underline{\zeta}}^{2}=0$. 

This implies that that all the eigenvalues have their extremal values
of $\zeta_{j}=\pm1$, which can only occur if there is a unitary transformation
that sends every mode into either a ground or excited state, which
is a pure state. As a consequence, we expect that $F=1$, since $F$
is just proportional to the linear entropy. This is exactly the result
given by the inner-product expression, Eq (\ref{eq:stretched variable overlap}).

Thus we see that the classical domain has a clear physical interpretation.
\emph{It is a domain whose boundary is the Gaussian pure states, with
the infinite temperature thermal state of highest entropy at the center.}

\subsection{Gaussian density operator Q-function}

For any Gaussian density operator, like a thermal or BCS state, such
that:
\begin{equation}
\hat{\rho}=\hat{\Lambda}(\underline{\underline{\sigma}}')=\hat{\Lambda}\left(\frac{1}{2}\left[\underline{\underline{\bar{I}}}-\underline{\underline{\zeta}}'\right]\right),
\end{equation}
 the fermionic Q-function in the limit of $k\rightarrow0$ has the
simple form:
\begin{equation}
Q\left(\underline{\underline{\zeta}}\right)=\frac{1}{\mathcal{V}}\det\left[\underline{\underline{I}}+\underline{\underline{\zeta}}'\underline{\underline{\zeta}}\right]^{1/2}.\label{eq:Gaussian Q-function}
\end{equation}
We have already given the simplest example of an infinite temperature
state with $\underline{\underline{\zeta}}'=0$. The corresponding
Q-function is simply the uniform Q-function, $Q=1/\mathcal{V}$. 

More generally, since any density matrix can be represented~\cite{Corney:2006_PRB}
using a positive fermionic P-function $P\left(\underline{\underline{\zeta}}'\right)$
in the stretched coordinates, then any density matrix has a Q-function
given by:
\begin{equation}
Q\left(\underline{\underline{\zeta}}\right)=\frac{1}{\mathcal{V}}\int d\underline{\underline{\zeta}}'P\left(\underline{\underline{\zeta}}'\right)\det\left[\underline{\underline{I}}+\underline{\underline{\zeta}}'\underline{\underline{\zeta}}\right]^{1/2}.
\end{equation}

A special behavior occurs in the case of a pure state, for which $P\left(\underline{\underline{\zeta}}\right)=\delta\left(\underline{\underline{\zeta}}-\underline{\underline{\zeta_{p}}}\right)$,
such that $\underline{\underline{\zeta_{p}^{2}}}=\underline{\underline{I}}$.
Let us call the corresponding pure state Q-function $Q_{p}$. Given
that it is a pure state $\underline{\underline{\zeta_{p}}}$, we may
wish to evaluate the Q-function at $\underline{\underline{\zeta}}=\underline{\underline{\zeta_{p}}}$,
which physically should be large from overlap arguments. The result
can be found using the Q-function definition of Eq (\ref{eq:Qf}),
together with the fact that a pure state density matrix is a projection
operator. Alternatively, one can use the Gaussian state result of
Eq (\ref{eq:Gaussian Q-function}). Either equation gives the same
result: the Q-function probability is increased by a factor of $2^{M}$
above its value in the highest entropy case:
\begin{equation}
Q_{p}\left(\underline{\underline{\zeta_{p}}}\right)=\frac{1}{\mathcal{N}}=\frac{2^{M}}{\mathcal{V}}\,.
\end{equation}
However, the Q-function is normalized to unity within the domain $\mathcal{V}$.
Therefore, the volume occupied by this high probability peak for a
pure state must be relatively small at large $M$, of order $2^{-M}\mathcal{V}$.

In the opposite extreme, we can always find an \emph{antipodean} state.
This is exactly on the opposite side of the classical domain boundary,
at $\underline{\underline{\zeta_{a}}}=-\underline{\underline{\zeta_{p}}}$.
We note that $\underline{\underline{\zeta_{a}}}\underline{\underline{\zeta_{p}}}=-\underline{\underline{\zeta_{p}^{2}}}=-\underline{\underline{I}}$.
Hence, $Q_{p}\left(\underline{\underline{\zeta_{a}}}\right)=0$ for
this pure state. That is, the Q-function of a pure state always vanishes
at the point antipodean to the original pure state. Physically, this
is caused by the fact that in the basis for which the pure state is
a number state, the antipodean state has every eigenvalue reflected,
and therefore it is orthogonal. 

There are many other orthogonal states to a pure state. In fact, there
are $2^{M}-1$ orthogonal, pure boundary states $\underline{\underline{\zeta_{o}}}$
of this type, in which one or more eigenvalues changes sign, leading
to a vanishing Q-function with $Q_{p}\left(\underline{\underline{\zeta_{o}}}\right)=0$.

We note that by changing the values of $k,s$, the detailed shape
of the Q-function can be modified. For example, if $s\gg0$ then the
distribution is always concentrated close to the origin, and the high
temperature state is a classical Gaussian similar to Eq (\ref{eq:Gaussian-integral}).
If $s\ll0$, the distribution is concentrated at the boundaries, and
gives a distribution over pure states.

\subsection{Single mode Gaussian operators}

The normal-ordered single-mode Gaussian operator~\cite{Corney:2006_JPA}
is a thermal density matrix with fermion number $n=1-\tilde{n}$:
\begin{eqnarray}
\hat{\Lambda}_{1}(n) & = & \tilde{n}:\exp\left[-\hat{a}^{\dagger}\hat{a}\left(2-\frac{1}{\tilde{n}}\right)\right]:.\nonumber \\
 & = & \left[\tilde{n}+\hat{a}^{\dagger}\hat{a}\left(2n-1\right)\right].\label{eq:GO_1M}
\end{eqnarray}
 Here we consider $n$ real and $n\in\left(0,1\right)$; $n$ gives
the number of particles, while $\tilde{n}=1-n$ gives the number of
holes in the thermal state that corresponds to $\hat{\Lambda}_{1}$.
As previously, it is useful to define a more symmetric variable, $\zeta=1-2n\in\left[-1,1\right]$.
Following the definition of the fermionic basis given in Eq. (\ref{eq:GaussianBasis}):
\begin{equation}
\hat{\Lambda}_{1}^{N}\left(\zeta\right)=\frac{1}{\mathcal{N}}\hat{\Lambda}_{1}\left(n\right)S_{1}\left(n\right),\label{eq:GOBasis_1M}
\end{equation}
where $S_{1}\left(n\right)$ is the corresponding normalization function
of Eq. (\ref{eq:Normfunction_P}) for $M=1$. If $s=0$, the normalization
function has a particle-hole symmetry, since it is an even function
of $\bar{n}$, given by:
\begin{equation}
S_{1}\left(\zeta^{2}\right)=\left(1-\zeta^{2}\right)^{k}=4^{k}\left(n\tilde{n}\right)^{k}.
\end{equation}

For the single mode case the resolution of unity for the phase-space
variable $n$ is given by:
\begin{equation}
\int_{-1}^{1}d\zeta\hat{\Lambda}_{1}^{N}\left(\zeta\right)=\hat{I}.\label{eq:ResUnity_1M}
\end{equation}
Here we give an explicit proof for resolution of unity for the special
case of one mode. Expanding the Gaussian, the integral over phase-space
is:
\begin{eqnarray}
\hat{I}_{G} & = & \frac{1}{2\mathcal{N}}\int_{-1}^{1}d\zeta\left[1+\zeta-2\zeta\hat{a}^{\dagger}\hat{a}\right]S_{1}\left(\zeta^{2}\right).
\end{eqnarray}
The odd-parity integral terms over $\bar{n}$ all vanish, including
the operator valued part. Hence, to prove a resolution of the identity,
it is only necessary to show that:
\begin{eqnarray}
1 & = & \frac{1}{\mathcal{N}}\int_{0}^{1}d\zeta\left(1-\zeta^{2}\right)^{k}.
\end{eqnarray}
A resolution of unity therefore requires a normalization of:
\begin{equation}
\mathcal{N}=\frac{\sqrt{\pi}k!}{2\Gamma\left[k+\frac{3}{2}\right]},
\end{equation}
in agreement with Eq (\ref{eq:normalization factor}). In the limit
of $k\rightarrow0$, one has $\mathcal{N}=\mathcal{V}/2=1$.

\subsection{Example of thermal states}

We will now give an explicit expression for the single-mode Q-function
described with this formalism. Since every single mode fermion state
is a thermal state, we need only consider the thermal states, with
the finite-temperature Fermi-Dirac mean occupation number of:
\begin{equation}
n_{th}=\frac{1}{\exp\left[\left(E-\mu\right)/k_{B}T\right]+1}\,.
\end{equation}
The density matrix is itself one of the Gaussian operators:
\begin{equation}
\hat{\rho}_{th}=\tilde{n}_{th}:\exp\left[-\hat{a}^{+}\left(2-\tilde{n}_{th}^{-1}\right)\hat{a}\right]:=\hat{\Lambda}_{th}\left(n_{th}\right)\:.\label{eq:DM-thermalStates}
\end{equation}

The explicit form of the fermionic Q-function for this case is given
by taking the inner product of Eq. (\ref{eq:DM-thermalStates}) and
the normalized Gaussian state $\hat{\Lambda}^{N}(\zeta)$. The symmetrized
thermal occupation is $\zeta_{th}=1-2n_{th}=2\tilde{n}_{th}-1$. Therefore,
in the $k,s\rightarrow0$ limit, the Q-function is given by:
\begin{eqnarray}
Q_{th}(\zeta) & = & {\rm Tr}\left[\hat{\Lambda}_{th}(n_{th})\hat{\Lambda}^{N}(\zeta)\right]\nonumber \\
 & = & \frac{1}{2{\cal N}}S_{1}\left(\zeta^{2}\right)\left(1+\zeta_{th}\zeta\right).\label{eq:QS-1M}
\end{eqnarray}
Here we have used the result for the trace of two normally ordered
normalized Gaussian operators~\cite{PRA_Entropy_paper_2011}. Since
$-1<\zeta,\zeta_{th}<1$, the $Q$ function is clearly positive, and
the result agrees with the general expression for a multi-mode Gaussian
Q-function, Eq (\ref{eq:Gaussian Q-function}).

\subsection{Q-function and observables }

Using the expression for the observables given in Eq. (\ref{eq:Moments_NO})
, we can write the corresponding differential identities as:
\begin{equation}
\hat{a}\hat{a}^{\dagger}\hat{\Lambda}_{1}=\left[\tilde{n}-n\tilde{n}\frac{\partial}{\partial n}\right]\hat{\Lambda}_{1}.
\end{equation}
Hence, since the normalization factor is $\mathcal{N}=1$, the observables
for the single mode case are:
\begin{eqnarray}
\left\langle \hat{a}\hat{a}^{\dagger}\right\rangle  & = & \int_{-1}^{1}d\zeta Q\left(\zeta\right)\left(\tilde{n}+\frac{\partial}{\partial n}\left[n\tilde{n}+\ln S_{1}\right]\right).\label{eq:Mom-1M}
\end{eqnarray}
On performing the derivatives, taking the limit of $k\rightarrow0$
and simplifying terms, we obtain:
\begin{eqnarray}
\left\langle \hat{a}\hat{a}^{\dagger}-\frac{1}{2}\right\rangle  & = & \frac{3}{2}\int_{-1}^{1}\zeta Q\left(\zeta\right)d\zeta.\label{eq:Obs-Q-1M}
\end{eqnarray}

Using the expression of the Q-function given in Eq. (\ref{eq:QS-1M})
and the expression for the observables given in Eq. (\ref{eq:Obs-Q-1M}),
we evaluate the observable, which is
\begin{eqnarray}
\left\langle 2\hat{a}\hat{a}^{\dagger}-1\right\rangle _{th} & = & 3\int_{-1}^{1}\zeta Q_{th}\left(\zeta\right)d\zeta\nonumber \\
 & = & \zeta_{th}.\label{eq:IntMoments-1M}
\end{eqnarray}
This result corresponds to the expected thermal average value.

\section{Summary\label{sec:Summary}}

In summary, we have introduced a probabilistic fermionic Q-function
for an arbitrary many body fermionic system. The fermionic Q-function
is physically interpreted as a suitably normalized overlap of the
density operator and the normally ordered Gaussian operators. 

We have also used three important properties of these Gaussian operators,
which are their positivity inside a bounded domain, their resolution
of unity and their differential identities. As a result, we have shown
that the Q-function is a continuous probability distribution, exists
for any quantum density matrix and has observables which are moments
of the distribution. 

The fermionic Q-function derived here is a general probabilistic approach
to fermion physics. It uses as a complete basis the Gaussian operators,
which have been applied in the study of many-body fermionic systems.
This is in contrast to previous fermionic Q-functions. For example,
a Q-function defined in terms of anticommuting Grassmann variables
is neither a probability nor even a real number. 

There is an elegant physical interpretation of the resulting phase-space
domain. These Gaussian states are physical states which have the highest
possible entropy at the domain centre, and the lowest possible entropy
at the boundaries. This is because the classical domain boundary is
the set of Gaussian pure states, which includes the BCS states. On
the other hand, its centre at $\underline{\underline{\zeta}}=0$ is
the infinite temperature thermal state of maximal entropy. 

Thus, as a hot fermionic system is cooled, we expect its Q-function
to be initially uniform on the classical domain, while gravitating
towards the boundary as it cools towards states of greater purity.
\begin{acknowledgments}
This research was supported by the Australian Research Council Discovery
Grants program. We would like to thank Terence Tao, Peter Forrester
and Joel Corney for useful discussions.
\end{acknowledgments}
\appendix

\section{Invariance of normal ordering under unitary transformations\label{sec:Invariance_PC}}

In this Appendix we prove that any unitary transformation of a class
D covariance matrix $\underline{\underline{\sigma}}$ that preserves
the class D symmetry, leaves the normal ordering form of the Gaussian
basis invariant. A similar result to this was obtained by Fan~\cite{Fan:2007}.
We note that this result is not generally true for an arbitrary unitary
transformation.

In these Appendices, we will use Grassmann algebra methods, and introduce
Grassmann coherent states defined as~\cite{Ohnuki_1978_CSFermiOp_PathInt,Cahill:1999}
$\left|\bm{\alpha}\right\rangle =\exp\left[\underline{\hat{a}}^{\dagger}\cdot\underline{\alpha}\right]\left|0\right\rangle $.
Here $\bm{\alpha}$ is an anti-commuting Grassmann variable with conjugate
$\overline{\bm{\alpha}}$, the eigenvalue equation is $\hat{\mathbf{a}}\left|\bm{\alpha}\right\rangle =\bm{\alpha}\left|\bm{\alpha}\right\rangle $,
and we use the following Grassmann coherent state identities,
\begin{eqnarray}
\int d^{2M}\bm{\alpha} & \left|\bm{\alpha}\right\rangle \left\langle \bm{\alpha}\right|= & \hat{I}\nonumber \\
{\rm Tr}\left(\hat{A}\right) & = & \int d^{2M}\bm{\alpha}\left\langle -\bm{\alpha}\right|\hat{A}\left|\bm{\alpha}\right\rangle 
\end{eqnarray}

\subsection*{Diagonalization transformation}

It is sufficient to prove this result for a diagonalization transformation,
since the matrices $\underline{\underline{\sigma}}$ are all hermitian
and therefore can always be made diagonal. For convenience, the covariance
matrix $\underline{\underline{\sigma}}$ is expressed in terms of
$\underline{\underline{\mu}}=\underline{\underline{\sigma}}^{-1}-2\underline{\underline{\bar{I}}}$.
We will show that on diagonalizing the matrix $\underline{\underline{\mu}}$,
the Gaussian operator
\begin{equation}
\hat{\Lambda}^{u}\left(\underline{\underline{\mu}}\right)=:\exp\left[-\hat{\underline{a}}^{\dagger}\underline{\underline{\mu}}\hat{\underline{a}}/2\right]:\,
\end{equation}
 remains in normally-ordered form, where the unitary transformation
$\underline{\underline{U}}$ that diagonalizes $\underline{\underline{\mu}}$
preserves the class D symmetry, and the unitarily transformed Fermi
operators are:
\begin{equation}
\hat{\underline{b}}=\underline{\underline{U}}\hat{\underline{a}}.\label{eq:Unitary_transf_operator-1}
\end{equation}
This implies that $\underline{\underline{\lambda}}=\underline{\underline{U}}\underline{\underline{\mu}}\underline{\underline{U}}^{-1}$
are the radial coordinates of the transformation and
\begin{equation}
\underline{\underline{\lambda}}=\left[\begin{array}{cc}
\bm{\lambda} & \bm{0}\\
\bm{0} & -\bm{\lambda}
\end{array}\right].
\end{equation}

To prove this result, we will use Grassmann coherent states. The following
double Grassmann integral is identical to the Gaussian operator $\hat{\Lambda}^{u}$:
\begin{equation}
\hat{I}_{\Lambda}=\int\int d\underline{\gamma}\left|\bm{\alpha}\right\rangle \left\langle \bm{\alpha}\right|\hat{\Lambda}^{u}\left(\underline{\underline{\mu}}\right)\left|\bm{\beta}\right\rangle \left\langle \bm{\beta}\right|,\label{eq:I1}
\end{equation}
where we have introduced the compact notation:
\begin{eqnarray}
\underline{\gamma} & = & \left(\begin{array}{c}
\bm{\beta}\\
\overline{\bm{\alpha}}^{T}
\end{array}\right),\qquad d\underline{\gamma}\equiv d^{2}\bm{\alpha}d^{2}\bm{\beta}.
\end{eqnarray}

\subsection*{Gaussian series expansion}

We wish to prove that $\hat{I}_{\Lambda}$, and hence $\hat{\Lambda}^{u}$,
is equal to a normally-ordered diagonal Gaussian form. To proceed,
$\hat{\Lambda}^{u}$ inside the integral is expanded as a series of
even order normally-ordered polynomials in the Fermi operators, so
that:
\begin{equation}
\hat{\Lambda}^{u}=\sum_{n=0}^{\infty}\frac{2^{-n}}{n!}P^{(2n)}\,.
\end{equation}

Using an explicit form of the matrix $\underline{\underline{\mu}}$
and recalling that $\underline{\hat{a}}=\left(\hat{\bm{a}},\hat{\bm{a}}^{\dagger}\right)$,
we can normally order the lowest order quadratic form $P^{(2)}$ of
$\hat{\Lambda}^{(u)}$ as:
\begin{eqnarray}
P^{(2)} & = & :\left(\hat{\bm{a}}^{\dagger},\hat{\bm{a}}\right)\left[\begin{array}{cc}
\bm{\mu} & \bm{\xi}\\
-\bm{\xi}^{*} & -\bm{\mu}^{T}
\end{array}\right]\left(\begin{array}{c}
\hat{\bm{a}}\\
\hat{\bm{a}}^{\dagger}
\end{array}\right):\nonumber \\
 & = & \hat{\bm{a}}^{\dagger}\bm{\mu}\hat{\bm{a}}+\hat{\bm{a}}^{\dagger}\bm{\xi}\hat{\bm{a}}^{\dagger}-\hat{\bm{a}}\bm{\xi}^{*}\hat{\bm{a}}+\hat{\bm{a}}^{\dagger}\bm{\mu}\hat{\bm{a}}.
\end{eqnarray}

In the last step there is a sign change as well as a transposition,
owing to the definition of fermionic normal-ordering, so that $:\hat{b}\hat{a}^{\dagger}:=-\hat{a}^{\dagger}\hat{b}$
. From the Grassmann coherent state eigenvalue properties and anticommutation
relations, we obtain:
\begin{eqnarray}
-\hat{\bm{a}}\bm{\xi}^{*}\hat{\bm{a}}\left|\bm{\beta}\right\rangle  & = & -\bm{\beta}^{T}\bm{\xi}^{*}\bm{\beta}\left|\bm{\beta}\right\rangle \nonumber \\
\left\langle \bm{\alpha}\right|\hat{\bm{a}}^{\dagger}\bm{\xi}\hat{\bm{a}}^{\dagger} & = & \left\langle \bm{\alpha}\right|\overline{\bm{\alpha}}\bm{\xi}\overline{\bm{\alpha}}^{T},
\end{eqnarray}
and applying this to all terms:
\begin{eqnarray}
\left\langle \bm{\alpha}\right|P^{(2)}\left(\underline{\hat{a}}\right)\left|\bm{\beta}\right\rangle  & = & \left\langle \bm{\alpha}\right.\left|\bm{\beta}\right\rangle \underline{\gamma}^{T}\underline{\underline{\Sigma}}\underline{\underline{\mu}}\underline{\gamma}\,,
\end{eqnarray}
where $\underline{\underline{\Sigma}}$ is the transposition matrix
defined in the main text.

For higher order polynomials, an analogous procedure occurs. There
will be an even number of sign changes for both Fermi operator re-orderings
and for Grassmann variable re-orderings. Hence:
\begin{eqnarray}
\left\langle \bm{\alpha}\right|\hat{\Lambda}^{u}\left|\bm{\beta}\right\rangle  & = & \exp\left[\frac{-1}{2}\underline{\gamma}^{T}\underline{\underline{\Sigma}}\underline{\underline{\mu}}\underline{\gamma}\right]\left\langle \bm{\alpha}\right.\left|\bm{\beta}\right\rangle .\label{eq:Grassmann-Gaussian-inner-product}
\end{eqnarray}

\subsection*{Unitary transformations}

So far we have expressed the un-normalized Gaussian operator term
in terms of Grassmann variables. Next, we are going to consider the
class-D unitary transformations that diagonalize the matrix $\underline{\underline{\mu}}$
under consideration. This is a $2M\times2M$ unitary transformation
of the Fermi operators:
\begin{equation}
\left(\begin{array}{c}
\hat{\bm{b}}\\
\hat{\bm{b}}^{\dagger}
\end{array}\right)=\underline{\underline{U}}\left(\begin{array}{c}
\hat{\bm{a}}\\
\hat{\bm{a}}^{\dagger}
\end{array}\right).\label{eq:Unitary_transf_operator-1-1}
\end{equation}
Note that if we diagonalize the matrix $\underline{\underline{\mu}}$
we also diagonalize the matrix $\underline{\underline{\sigma}}$.
As pointed out by Altland and Zirnbauer~\cite{Altland_Zirnbauer:1997},
$\underline{\underline{U}}$ must satisfy $\underline{\underline{U}}=\underline{\underline{\Sigma}}\underline{\underline{U}}^{*}\underline{\underline{\Sigma}}\,$
with: 
\begin{eqnarray}
\underline{\underline{U}} & = & \left[\begin{array}{cc}
\mathbf{u}^{11} & \mathbf{u}^{12}\\
\mathbf{u}^{21} & \mathbf{u}^{22}
\end{array}\right]\,.
\end{eqnarray}

This property implies that:
\begin{eqnarray}
\left[\begin{array}{cc}
\mathbf{u}^{11} & \mathbf{u}^{12}\\
\mathbf{u}^{21} & \mathbf{u}^{22}
\end{array}\right] & = & \left[\begin{array}{cc}
\mathbf{u}^{22*} & \mathbf{u}^{21*}\\
\mathbf{u}^{12*} & \mathbf{u}^{11*}
\end{array}\right],
\end{eqnarray}
which guarantees that the transformation, when applied to the Fermi
operators, preserves operator commutators~\cite{Balian:1969}. We
can apply the same transformation to the Grassmann variables, by choosing
that 
\begin{eqnarray}
\left(\begin{array}{c}
\bm{\beta}'\\
\overline{\bm{\alpha}}'
\end{array}\right) & = & \underline{\underline{U}}\left(\begin{array}{c}
\bm{\beta}\\
\overline{\bm{\alpha}}
\end{array}\right)\nonumber \\
\left(\begin{array}{c}
\bm{\alpha}'\\
\overline{\bm{\beta}}'
\end{array}\right) & = & \underline{\underline{U}}\left(\begin{array}{c}
\bm{\alpha}\\
\overline{\bm{\beta}}
\end{array}\right)\,.
\end{eqnarray}

From the first expression, one obtains:
\begin{equation}
\overline{\bm{\alpha}}'=\mathbf{u}^{21}\bm{\beta}+\mathbf{u}^{22}\overline{\bm{\alpha}}\,,
\end{equation}
while on conjugating the second expression:
\begin{equation}
\bm{\alpha}'^{*}=\mathbf{u}^{12*}\bm{\beta}+\mathbf{u}^{11*}\overline{\bm{\alpha}}=\mathbf{u}^{21}\bm{\beta}+\mathbf{u}^{22}\overline{\bm{\alpha}}=\overline{\bm{\alpha}}'\,.
\end{equation}

This means that the transformation is a consistent transformation
on Grassmann variables that preserves conjugation, so that if $\left|\bm{\alpha}'\right\rangle '$
is the new coherent state in the new basis, then:
\begin{equation}
\left|\bm{\alpha}\right\rangle =\left|\bm{\alpha}'\right\rangle '\,.
\end{equation}
Therefore, the Grassmann quadratic in the exponential is also diagonalized
by the unitary transformation, so that:
\begin{equation}
\underline{\gamma}^{T}\underline{\underline{\Sigma}}\underline{\underline{\mu}}\underline{\gamma}=\underline{\gamma}^{\prime T}\underline{\underline{\Sigma}}\underline{\underline{\lambda}}\underline{\gamma}^{\prime}=-2\overline{\bm{\alpha}}'\bm{\lambda}\bm{\beta}'.
\end{equation}
 It follows that one can extend this argument to each even order polynomial
form in the exponential, so that: 
\begin{equation}
\exp\left[-\underline{\gamma}^{T}\underline{\underline{\Sigma}}\underline{\underline{\mu}}\underline{\gamma}/2\right]=\exp\left[-\underline{\gamma}^{\prime T}\underline{\underline{\Sigma}}\underline{\underline{\lambda}}\underline{\gamma}^{\prime}/2\right]\,.
\end{equation}

\subsection*{Gaussian in operator form}

We can now evaluate the Grassmann integral, noting that the integration
measure is invariant under a unitary transformation, since it has
unit Jacobian:
\begin{eqnarray}
\hat{I}_{\Lambda} & = & \int\int d\underline{\gamma}\left|\bm{\alpha}\right\rangle \left\langle \bm{\alpha}\right|\exp\left[-\underline{\gamma}^{T}\underline{\underline{\Sigma}}\underline{\underline{\mu}}\underline{\gamma}/2\right]\left|\bm{\beta}\right\rangle \left\langle \bm{\beta}\right|\nonumber \\
 & = & \int\int d\underline{\gamma^{\prime}}\left|\bm{\alpha}'\right\rangle '\left\langle \bm{\alpha}'\right|'\exp\left[-\underline{\gamma}^{\prime T}\underline{\underline{\Sigma}}\underline{\underline{\lambda}}\underline{\gamma}^{\prime}/2\right]\left|\bm{\beta}'\right\rangle '\left\langle \bm{\beta}'\right|'.\nonumber \\
 &  & \,
\end{eqnarray}

We can now use the eigenvalue properties of the Grassmann coherent
states in the transformed basis to return back to the operator form
in the diagonal basis, so that, on making use of the Grassmann coherent
state expansion of the identity, one obtains: 
\begin{eqnarray}
\hat{\Lambda}^{(u)} & = & \int\int d\underline{\gamma^{\prime}}\left|\bm{\alpha}^{\prime}\right\rangle ^{\prime}\left\langle \bm{\alpha}'\right|^{\prime}:\exp\left[-\hat{\underline{b}}^{\dagger}\underline{\underline{\lambda}}\hat{\underline{b}}/2\right]:\left|\bm{\beta}'\right\rangle '\left\langle \bm{\beta}'\right|'\nonumber \\
 & = & :\exp\left[-\hat{\underline{b}}^{\dagger}\underline{\underline{\lambda}}\hat{\underline{b}}/2\right]:\,.
\end{eqnarray}
This means that we have the required result. That is, we can diagonalize
the class-D covariance matrix for the normally-ordered Gaussian operators
using the same transformations used in~\cite{ResUnityFGO:2013},
while leaving the underlying normally-ordered form invariant, just
as one can in a non-normally-ordered case.

\section{Inner product of two Gaussian operators\label{sec:Appendix-B:-Inner} }

Consider two normalized Gaussian operators, $\hat{\Lambda}(\underline{\underline{\sigma}})$
and $\hat{\Lambda}(\underline{\underline{\sigma}}')$. We wish to
evaluate their standard Hilbert-Schmidt inner product:
\begin{equation}
F\left(\underline{\underline{\sigma}},\underline{\underline{\sigma}}'\right)={\rm Tr}\left(\hat{\Lambda}(\underline{\underline{\sigma}})\hat{\Lambda}(\underline{\underline{\sigma}}')\right).
\end{equation}

We will start be evaluating the inner product of the two corresponding
un-normalized operators, $\hat{\Lambda}^{u}\left(\underline{\underline{\mu}}\right)$
and $\hat{\Lambda}^{u}(\underline{\underline{\mu}}')$, where:
\begin{eqnarray}
F^{u}\left(\underline{\underline{\mu}},\underline{\underline{\mu}}'\right) & \equiv & {\rm Tr}\left(\hat{\Lambda}^{u}(\underline{\underline{\mu}})\hat{\Lambda}^{u}(\underline{\underline{\mu}}')\right).
\end{eqnarray}
We now use the known result for a trace, and the expansion of the
identity operator in Grassmann variables, so that:

\begin{eqnarray}
F^{u}\left(\underline{\underline{\mu}},\underline{\underline{\mu}}'\right) & = & \int d^{2M}\underline{\alpha}\left\langle -\bm{\alpha}\right|\hat{\Lambda}^{u}(\underline{\underline{\mu}})\hat{\Lambda}^{u}(\underline{\underline{\mu}}')\left|\bm{\alpha}\right\rangle \\
 & = & \int d^{2M}\underline{\alpha}d^{2M}\underline{\beta}\left\langle -\bm{\alpha}\right|\hat{\Lambda}^{u}(\underline{\underline{\mu}})\left|\bm{\beta}\right\rangle \left\langle \bm{\beta}\right|\hat{\Lambda}^{u}(\underline{\underline{\mu}}')\left|\bm{\alpha}\right\rangle .\nonumber 
\end{eqnarray}

From the inner product properties of the Grassmann coherent states,
\begin{equation}
\left\langle \bm{\beta}\right|\left|\bm{\alpha}\right\rangle =e^{\left[\bar{\bm{\beta}}\bm{\alpha}-\left(\bar{\bm{\alpha}}\bm{\alpha}+\bar{\bm{\beta}}\bm{\beta}\right)/2\right]}.
\end{equation}
Now, on taking account of normal ordering, together with the eigenvalue
properties of the coherent states, one can write that:
\begin{eqnarray}
\left\langle \bm{\beta}\right|\hat{\Lambda}^{u}(\underline{\underline{\mu}}')\left|\bm{\alpha}\right\rangle  & = & \left\langle \bm{\beta}\right|\left|\bm{\alpha}\right\rangle \exp^{-\left[\bar{\bm{\beta}},\bm{\alpha}\right]\underline{\underline{\mu}}'\left[\begin{array}{c}
\bm{\alpha}\\
\bar{\bm{\beta}}
\end{array}\right]/2}\\
 & = & e^{-\left\{ \left[\bar{\bm{\beta}},\bm{\alpha}\right]\left(\underline{\underline{\mu}}'+\bar{\underline{\underline{I}}}\right)\left[\begin{array}{c}
\bm{\alpha}\\
\bar{\bm{\beta}}
\end{array}\right]+\bar{\bm{\alpha}}\bm{\alpha}+\bar{\bm{\beta}}\bm{\beta}\right\} /2}.\nonumber 
\end{eqnarray}

Similarly, for the other term, 
\begin{eqnarray}
\left\langle -\bm{\alpha}\right|\hat{\Lambda}^{u}(\underline{\underline{\mu}})\left|\bm{\beta}\right\rangle  & = & \left\langle -\bm{\alpha}\right|\left|\bm{\beta}\right\rangle e^{-\left[-\bar{\bm{\alpha}},\bm{\beta}\right]\underline{\underline{\mu}}\left[\begin{array}{c}
\bm{\beta}\\
-\bar{\bm{\alpha}}
\end{array}\right]/2}\\
 & = & e^{-\left\{ \left[-\bar{\bm{\alpha}},\bm{\beta}\right]\left(\underline{\underline{\mu}}+\bar{\underline{\underline{I}}}\right)\left[\begin{array}{c}
\bm{\beta}\\
-\bar{\bm{\alpha}}
\end{array}\right]-\bar{\bm{\alpha}}\bm{\alpha}+\bar{\bm{\beta}}\bm{\beta}\right\} /2}.\nonumber 
\end{eqnarray}

\subsection*{Sign reversal variable change}

It is convenient at this stage to make a variable change of $\bar{\bm{\alpha}}=-\bar{\bm{\alpha}}'$,
noting that in Grassmann calculus, the conjugate variable can be regarded
as an independent complex variable. This causes a sign change in the
integration of $\left(-1\right)^{M}$. As a result, we can introduce
a new variable, 
\begin{equation}
\underline{\gamma}=\left[\begin{array}{c}
\bm{\alpha}\\
\bar{\bm{\beta}}\\
\bm{\beta}\\
\bar{\bm{\alpha}}'
\end{array}\right],
\end{equation}
and its conjugate, which is: 
\begin{equation}
\underline{\bar{\gamma}}=\left[\bar{\bm{\alpha}}',\bm{\beta},\bar{\bm{\beta}},\bm{\alpha}\right].
\end{equation}

We note that $d^{2M}\underline{\alpha}d^{2M}\underline{\beta}=\left(-1\right)^{M}d^{2M}\underline{\alpha}d^{2M}\underline{\bar{\beta}}=d^{4M}\underline{\gamma}$,
since the sign change and the Grassmann re-ordering in the integration
variables will cancel each other. As a result, we can write that 
\begin{equation}
F^{u}\left(\underline{\underline{\mu}},\underline{\underline{\nu}}\right)=\int d^{4M}\underline{\gamma}\exp\left[\frac{-1}{2}\underline{\bar{\gamma}}\underline{\underline{\Gamma}}\underline{\gamma}\right],
\end{equation}
where $\underline{\underline{\Gamma}}$ is a $4M\times4M$ matrix,
such that: 
\begin{equation}
\underline{\underline{\Gamma}}=\left[\begin{array}{cc}
\underline{\underline{I}} & \bar{\underline{\underline{I}}}+\underline{\underline{\mu}}\\
\bar{\underline{\underline{I}}}+\underline{\underline{\mu}}' & -\underline{\underline{I}}
\end{array}\right].
\end{equation}

\subsection*{Grassmann integration result}

This is a standard Grassmann gaussian integral, except in a space
of twice the previous dimension. Hence, the inner product is a Pfaffian
of the antisymmetrized form of $\underline{\underline{\Gamma}}$,
given as determinant by:
\begin{equation}
F^{u}=\det\left[\left(\bar{\underline{\underline{I}}}+\underline{\underline{\mu}}'\right)\left(\bar{\underline{\underline{I}}}+\underline{\underline{\mu}}\right)+\bar{\underline{\underline{I}}}^{2}\right]^{1/2}.
\end{equation}

Now, changing variables to the covariance matrix, one obtains: 
\begin{equation}
\bar{\underline{\underline{I}}}+\underline{\underline{\mu}}=\underline{\underline{\sigma}}^{-1}-\bar{\underline{\underline{I}}},
\end{equation}
so that in terms of these variables,
\[
F^{u}=\det\left[\left(\underline{\underline{\sigma}}'^{-1}-\bar{\underline{\underline{I}}}\right)\left(\underline{\underline{\sigma}}^{-1}-\bar{\underline{\underline{I}}}\right)+\bar{\underline{\underline{I}}}^{2}\right]^{1/2}
\]

Hence, multiplying by the normalizing factors of $\sqrt{\det\left[i\underline{\underline{\sigma}}\right]}$,
and noting that $\left(-1\right)^{2M}=1$, one obtains: 
\begin{eqnarray}
F & = & \det\left[\left(\bar{\underline{\underline{I}}}-\underline{\underline{\sigma}}'\right)\left(\bar{\underline{\underline{I}}}-\underline{\underline{\sigma}}\right)+\underline{\underline{\sigma}}'\underline{\underline{\sigma}}\right]^{1/2}.
\end{eqnarray}

This can also be written in terms of the stretched variables, $\underline{\underline{\zeta}}=\underline{\underline{\bar{I}}}-2\underline{\underline{\sigma}}$.
We note that:
\begin{eqnarray*}
\left(\bar{\underline{\underline{I}}}-\underline{\underline{\sigma}}'\right)\left(\bar{\underline{\underline{I}}}-\underline{\underline{\sigma}}\right) & = & \frac{1}{4}\left(\bar{\underline{\underline{I}}}+\underline{\underline{\zeta}}'\right)\left(\bar{\underline{\underline{I}}}+\underline{\underline{\zeta}}\right)\\
\underline{\underline{\sigma}}'\underline{\underline{\sigma}} & = & \frac{1}{4}\left(\bar{\underline{\underline{I}}}-\underline{\underline{\zeta}}'\right)\left(\bar{\underline{\underline{I}}}-\underline{\underline{\zeta}}\right)
\end{eqnarray*}
to give the result that:
\begin{eqnarray}
F & =2^{-M} & \sqrt{\det\left[\underline{\underline{I}}+\underline{\underline{\zeta}}'\underline{\underline{\zeta}}\right]}.
\end{eqnarray}

\bibliographystyle{apsrev4-1}

\begin{thebibliography}{48}%
\makeatletter
\providecommand \@ifxundefined [1]{%
 \@ifx{#1\undefined}
}%
\providecommand \@ifnum [1]{%
 \ifnum #1\expandafter \@firstoftwo
 \else \expandafter \@secondoftwo
 \fi
}%
\providecommand \@ifx [1]{%
 \ifx #1\expandafter \@firstoftwo
 \else \expandafter \@secondoftwo
 \fi
}%
\providecommand \natexlab [1]{#1}%
\providecommand \enquote  [1]{``#1''}%
\providecommand \bibnamefont  [1]{#1}%
\providecommand \bibfnamefont [1]{#1}%
\providecommand \citenamefont [1]{#1}%
\providecommand \href@noop [0]{\@secondoftwo}%
\providecommand \href [0]{\begingroup \@sanitize@url \@href}%
\providecommand \@href[1]{\@@startlink{#1}\@@href}%
\providecommand \@@href[1]{\endgroup#1\@@endlink}%
\providecommand \@sanitize@url [0]{\catcode `\\12\catcode `\$12\catcode
  `\&12\catcode `\#12\catcode `\^12\catcode `\_12\catcode `\%12\relax}%
\providecommand \@@startlink[1]{}%
\providecommand \@@endlink[0]{}%
\providecommand \url  [0]{\begingroup\@sanitize@url \@url }%
\providecommand \@url [1]{\endgroup\@href {#1}{\urlprefix }}%
\providecommand \urlprefix  [0]{URL }%
\providecommand \Eprint [0]{\href }%
\@ifxundefined \urlstyle {%
  \providecommand \doi  [0]{\begingroup \@sanitize@url \@doi}%
  \providecommand \@doi [1]{\endgroup \@@startlink {\doibase
  #1}doi:\discretionary {}{}{}#1\@@endlink }%
}{%
  \providecommand \doi  [0]{doi:\discretionary{}{}{}\begingroup
  \urlstyle{rm}\Url }%
}%
\providecommand \doibase [0]{http://dx.doi.org/}%
\providecommand \Doi [0]{\begingroup \@sanitize@url \@Doi }%
\providecommand \@Doi  [1]{\endgroup\@@startlink{\doibase#1}\@@Doi}%
\providecommand \@@Doi [1]{#1\@@endlink}%
\providecommand \selectlanguage [0]{\@gobble}%
\providecommand \bibinfo  [0]{\@secondoftwo}%
\providecommand \bibfield  [0]{\@secondoftwo}%
\providecommand \translation [1]{[#1]}%
\providecommand \BibitemOpen [0]{}%
\providecommand \bibitemStop [0]{}%
\providecommand \bibitemNoStop [0]{.\EOS\space}%
\providecommand \EOS [0]{\spacefactor3000\relax}%
\providecommand \BibitemShut  [1]{\csname bibitem#1\endcsname}%
\bibitem [{\citenamefont {Wigner}(1932)}]{Wigner_1932}%
  \BibitemOpen
  \bibfield  {author} {\bibinfo {author} {\bibfnamefont {E.}~\bibnamefont
  {Wigner}},\ }\href@noop {} {\bibfield  {journal} {\bibinfo  {journal} {Phys.
  Rev.},\ }\textbf {\bibinfo {volume} {40}},\ \bibinfo {pages} {749} (\bibinfo
  {year} {1932})}\BibitemShut {NoStop}%
\bibitem [{\citenamefont {Moyal}(1949)}]{Moyal_1949}%
  \BibitemOpen
  \bibfield  {author} {\bibinfo {author} {\bibfnamefont {J.~E.}\ \bibnamefont
  {Moyal}},\ }\href@noop {} {\bibfield  {journal} {\bibinfo  {journal} {Proc.
  Cambridge Phil. Soc.},\ }\textbf {\bibinfo {volume} {45}},\ \bibinfo {pages}
  {99} (\bibinfo {year} {1949})}\BibitemShut {NoStop}%
\bibitem [{\citenamefont {Glauber}(1963){\natexlab{a}}}]{Glauber:1963}%
  \BibitemOpen
  \bibfield  {author} {\bibinfo {author} {\bibfnamefont {R.~J.}\ \bibnamefont
  {Glauber}},\ }\href@noop {} {\bibfield  {journal} {\bibinfo  {journal} {Phys.
  Rev.},\ }\textbf {\bibinfo {volume} {130}},\ \bibinfo {pages} {2529}
  (\bibinfo {year} {1963}{\natexlab{a}})}\BibitemShut {NoStop}%
\bibitem [{\citenamefont {Husimi}(1940)}]{Husimi1940}%
  \BibitemOpen
  \bibfield  {author} {\bibinfo {author} {\bibfnamefont {K.}~\bibnamefont
  {Husimi}},\ }\href@noop {} {\bibfield  {journal} {\bibinfo  {journal} {Proc.
  Phys. Math. Soc. Jpn},\ }\textbf {\bibinfo {volume} {22}},\ \bibinfo {pages}
  {264} (\bibinfo {year} {1940})}\BibitemShut {NoStop}%
\bibitem [{\citenamefont {Arecchi}\ \emph {et~al.}(1972)\citenamefont
  {Arecchi}, \citenamefont {Courtens}, \citenamefont {Gilmore},\ and\
  \citenamefont {Thomas}}]{Arecchi:1972}%
  \BibitemOpen
  \bibfield  {author} {\bibinfo {author} {\bibfnamefont {F.~T.}\ \bibnamefont
  {Arecchi}}, \bibinfo {author} {\bibfnamefont {E.}~\bibnamefont {Courtens}},
  \bibinfo {author} {\bibfnamefont {R.}~\bibnamefont {Gilmore}}, \ and\
  \bibinfo {author} {\bibfnamefont {H.}~\bibnamefont {Thomas}},\ }\href@noop {}
  {\bibfield  {journal} {\bibinfo  {journal} {Phys. Rev. A},\ }\textbf
  {\bibinfo {volume} {6}},\ \bibinfo {pages} {2211} (\bibinfo {year}
  {1972})}\BibitemShut {NoStop}%
\bibitem [{\citenamefont {Gilmore}\ \emph {et~al.}(1975)\citenamefont
  {Gilmore}, \citenamefont {Bowden},\ and\ \citenamefont
  {Narducci}}]{Gilmore1975}%
  \BibitemOpen
  \bibfield  {author} {\bibinfo {author} {\bibfnamefont {R.}~\bibnamefont
  {Gilmore}}, \bibinfo {author} {\bibfnamefont {C.}~\bibnamefont {Bowden}}, \
  and\ \bibinfo {author} {\bibfnamefont {L.}~\bibnamefont {Narducci}},\ }\Doi
  {10.1103/PhysRevA.12.1019} {\bibfield  {journal} {\bibinfo  {journal} {Phys.
  Rev. A},\ }\textbf {\bibinfo {volume} {12}},\ \bibinfo {pages} {1019}
  (\bibinfo {year} {1975})}\BibitemShut {NoStop}%
\bibitem [{\citenamefont {Drummond}\ and\ \citenamefont
  {Gardiner}(1980)}]{Drummond:1980}%
  \BibitemOpen
  \bibfield  {author} {\bibinfo {author} {\bibfnamefont {P.~D.}\ \bibnamefont
  {Drummond}}\ and\ \bibinfo {author} {\bibfnamefont {C.~W.}\ \bibnamefont
  {Gardiner}},\ }\href@noop {} {\bibfield  {journal} {\bibinfo  {journal} {J.
  Phys. A},\ }\textbf {\bibinfo {volume} {13}},\ \bibinfo {pages} {2353}
  (\bibinfo {year} {1980})}\BibitemShut {NoStop}%
\bibitem [{\citenamefont {Corney}\ and\ \citenamefont
  {Drummond}(2003)}]{Corney:2003}%
  \BibitemOpen
  \bibfield  {author} {\bibinfo {author} {\bibfnamefont {J.~F.}\ \bibnamefont
  {Corney}}\ and\ \bibinfo {author} {\bibfnamefont {P.~D.}\ \bibnamefont
  {Drummond}},\ }\href@noop {} {\bibfield  {journal} {\bibinfo  {journal}
  {Phys. Rev. A},\ }\textbf {\bibinfo {volume} {68}},\ \bibinfo {pages}
  {063822} (\bibinfo {year} {2003})}\BibitemShut {NoStop}%
\bibitem [{\citenamefont {Corney}\ and\ \citenamefont
  {Drummond}(2004)}]{Corney:2004}%
  \BibitemOpen
  \bibfield  {author} {\bibinfo {author} {\bibfnamefont {J.~F.}\ \bibnamefont
  {Corney}}\ and\ \bibinfo {author} {\bibfnamefont {P.~D.}\ \bibnamefont
  {Drummond}},\ }\href@noop {} {\bibfield  {journal} {\bibinfo  {journal}
  {Phys. Rev. Lett.},\ }\textbf {\bibinfo {volume} {93}},\ \bibinfo {pages}
  {260401} (\bibinfo {year} {2004})}\BibitemShut {NoStop}%
\bibitem [{\citenamefont {Corney}\ and\ \citenamefont
  {Drummond}(2006){\natexlab{a}}}]{Corney:2006_PRB}%
  \BibitemOpen
  \bibfield  {author} {\bibinfo {author} {\bibfnamefont {J.~F.}\ \bibnamefont
  {Corney}}\ and\ \bibinfo {author} {\bibfnamefont {P.~D.}\ \bibnamefont
  {Drummond}},\ }\href@noop {} {\bibfield  {journal} {\bibinfo  {journal}
  {Phys. Rev. B},\ }\textbf {\bibinfo {volume} {73}},\ \bibinfo {pages}
  {125112} (\bibinfo {year} {2006}{\natexlab{a}})}\BibitemShut {NoStop}%
\bibitem [{\citenamefont {Reid}\ \emph {et~al.}(2014)\citenamefont {Reid},
  \citenamefont {Opanchuk}, \citenamefont {Rosales-Z\'arate},\ and\
  \citenamefont {Drummond}}]{Reid2014Multi}%
  \BibitemOpen
  \bibfield  {author} {\bibinfo {author} {\bibfnamefont {M.~D.}\ \bibnamefont
  {Reid}}, \bibinfo {author} {\bibfnamefont {B.}~\bibnamefont {Opanchuk}},
  \bibinfo {author} {\bibfnamefont {L.}~\bibnamefont {Rosales-Z\'arate}}, \
  and\ \bibinfo {author} {\bibfnamefont {P.~D.}\ \bibnamefont {Drummond}},\
  }\Doi {10.1103/PhysRevA.90.012111} {\bibfield  {journal} {\bibinfo  {journal}
  {Phys. Rev. A},\ }\textbf {\bibinfo {volume} {90}},\ \bibinfo {pages}
  {012111} (\bibinfo {year} {2014})}\BibitemShut {NoStop}%
\bibitem [{\citenamefont {Haake}\ \emph {et~al.}(1979)\citenamefont {Haake},
  \citenamefont {King}, \citenamefont {Schr{\"o}der}, \citenamefont {Haus},\
  and\ \citenamefont {Glauber}}]{Haake:1979}%
  \BibitemOpen
  \bibfield  {author} {\bibinfo {author} {\bibfnamefont {F.}~\bibnamefont
  {Haake}}, \bibinfo {author} {\bibfnamefont {H.}~\bibnamefont {King}},
  \bibinfo {author} {\bibfnamefont {G.}~\bibnamefont {Schr{\"o}der}}, \bibinfo
  {author} {\bibfnamefont {J.}~\bibnamefont {Haus}}, \ and\ \bibinfo {author}
  {\bibfnamefont {R.}~\bibnamefont {Glauber}},\ }\href@noop {} {\bibfield
  {journal} {\bibinfo  {journal} {Phys. Rev. A},\ }\textbf {\bibinfo {volume}
  {20}},\ \bibinfo {pages} {2047} (\bibinfo {year} {1979})}\BibitemShut
  {NoStop}%
\bibitem [{\citenamefont {Lee}(1984)}]{Lee:1984}%
  \BibitemOpen
  \bibfield  {author} {\bibinfo {author} {\bibfnamefont {C.~T.}\ \bibnamefont
  {Lee}},\ }\Doi {10.1103/PhysRevA.30.3308} {\bibfield  {journal} {\bibinfo
  {journal} {Phys. Rev. A},\ }\textbf {\bibinfo {volume} {30}},\ \bibinfo
  {pages} {3308} (\bibinfo {year} {1984})}\BibitemShut {NoStop}%
\bibitem [{\citenamefont {Drummond}(1984)}]{Drummond1984PhysLetts}%
  \BibitemOpen
  \bibfield  {author} {\bibinfo {author} {\bibfnamefont {P.~D.}\ \bibnamefont
  {Drummond}},\ }\href@noop {} {\bibfield  {journal} {\bibinfo  {journal}
  {Phys. Letts.},\ }\textbf {\bibinfo {volume} {106A}},\ \bibinfo {pages} {118}
  (\bibinfo {year} {1984})}\BibitemShut {NoStop}%
\bibitem [{\citenamefont {Altland}\ and\ \citenamefont
  {Haake}(2012){\natexlab{a}}}]{Altland:2012_NJP}%
  \BibitemOpen
  \bibfield  {author} {\bibinfo {author} {\bibfnamefont {A.}~\bibnamefont
  {Altland}}\ and\ \bibinfo {author} {\bibfnamefont {F.}~\bibnamefont
  {Haake}},\ }\href {http://stacks.iop.org/1367-2630/14/i=7/a=073011}
  {\bibfield  {journal} {\bibinfo  {journal} {New Journal of Physics},\
  }\textbf {\bibinfo {volume} {14}},\ \bibinfo {pages} {073011} (\bibinfo
  {year} {2012}{\natexlab{a}})}\BibitemShut {NoStop}%
\bibitem [{\citenamefont {Altland}\ and\ \citenamefont
  {Haake}(2012){\natexlab{b}}}]{Altland_PRL2012_Qchaos}%
  \BibitemOpen
  \bibfield  {author} {\bibinfo {author} {\bibfnamefont {A.}~\bibnamefont
  {Altland}}\ and\ \bibinfo {author} {\bibfnamefont {F.}~\bibnamefont
  {Haake}},\ }\href@noop {} {\bibfield  {journal} {\bibinfo  {journal} {Phys.
  Rev. Lett.},\ }\textbf {\bibinfo {volume} {108}},\ \bibinfo {pages} {073601}
  (\bibinfo {year} {2012}{\natexlab{b}})}\BibitemShut {NoStop}%
\bibitem [{\citenamefont {Corney}\ and\ \citenamefont
  {Drummond}(2006){\natexlab{b}}}]{Corney:2006_JPA}%
  \BibitemOpen
  \bibfield  {author} {\bibinfo {author} {\bibfnamefont {J.~F.}\ \bibnamefont
  {Corney}}\ and\ \bibinfo {author} {\bibfnamefont {P.~D.}\ \bibnamefont
  {Drummond}},\ }\href@noop {} {\bibfield  {journal} {\bibinfo  {journal} {J.
  Phys. A},\ }\textbf {\bibinfo {volume} {39}},\ \bibinfo {pages} {269}
  (\bibinfo {year} {2006}{\natexlab{b}})}\BibitemShut {NoStop}%
\bibitem [{\citenamefont {Rosales-Z\'arate}\ and\ \citenamefont
  {Drummond}(2013)}]{ResUnityFGO:2013}%
  \BibitemOpen
  \bibfield  {author} {\bibinfo {author} {\bibfnamefont {L.~E.~C.}\
  \bibnamefont {Rosales-Z\'arate}}\ and\ \bibinfo {author} {\bibfnamefont
  {P.~D.}\ \bibnamefont {Drummond}},\ }\href
  {http://stacks.iop.org/1751-8121/46/i=27/a=275203} {\bibfield  {journal}
  {\bibinfo  {journal} {J. Phys. A},\ }\textbf {\bibinfo {volume} {46}},\
  \bibinfo {pages} {275203} (\bibinfo {year} {2013})}\BibitemShut {NoStop}%
\bibitem [{\citenamefont {Altland}\ and\ \citenamefont
  {Zirnbauer}(1997)}]{Altland_Zirnbauer:1997}%
  \BibitemOpen
  \bibfield  {author} {\bibinfo {author} {\bibfnamefont {A.}~\bibnamefont
  {Altland}}\ and\ \bibinfo {author} {\bibfnamefont {M.~R.}\ \bibnamefont
  {Zirnbauer}},\ }\href@noop {} {\bibfield  {journal} {\bibinfo  {journal}
  {Phys. Rev. B},\ }\textbf {\bibinfo {volume} {55}},\ \bibinfo {pages} {1142}
  (\bibinfo {year} {1997})}\BibitemShut {NoStop}%
\bibitem [{\citenamefont {Perelomov}(1986)}]{Perelomov_book_Coherent_state}%
  \BibitemOpen
  \bibfield  {author} {\bibinfo {author} {\bibfnamefont {A.~M.}\ \bibnamefont
  {Perelomov}},\ }\href@noop {} {\emph {\bibinfo {title} {Generalized coherent
  states and their applications}}},\ Texts and monographs in physics\ (\bibinfo
   {publisher} {Springer, Berlin},\ \bibinfo {year} {1986})\BibitemShut
  {NoStop}%
\bibitem [{\citenamefont {Stratonovich}(1956)}]{Stratonovich1956}%
  \BibitemOpen
  \bibfield  {author} {\bibinfo {author} {\bibfnamefont {R.~L.}\ \bibnamefont
  {Stratonovich}},\ }\href@noop {} {\bibfield  {journal} {\bibinfo  {journal}
  {Sov. Phys. JETP},\ }\textbf {\bibinfo {volume} {31}},\ \bibinfo {pages}
  {1012} (\bibinfo {year} {1956})}\BibitemShut {NoStop}%
\bibitem [{\citenamefont {Weyl}(1939)}]{Weyl_book_groups}%
  \BibitemOpen
  \bibfield  {author} {\bibinfo {author} {\bibfnamefont {H.}~\bibnamefont
  {Weyl}},\ }\href@noop {} {\emph {\bibinfo {title} {{The classical groups:
  Their invariants and representations}}}}\ (\bibinfo  {publisher} {Princeton
  University Press},\ \bibinfo {address} {Princeton, N.J.},\ \bibinfo {year}
  {1939})\BibitemShut {NoStop}%
\bibitem [{\citenamefont {Aimi}\ and\ \citenamefont
  {Imada}(2007)}]{Imada:2007_GBMC}%
  \BibitemOpen
  \bibfield  {author} {\bibinfo {author} {\bibfnamefont {T.}~\bibnamefont
  {Aimi}}\ and\ \bibinfo {author} {\bibfnamefont {M.}~\bibnamefont {Imada}},\
  }\href@noop {} {\bibfield  {journal} {\bibinfo  {journal} {J. Phys. Soc.
  Jpn.},\ }\textbf {\bibinfo {volume} {76}},\ \bibinfo {pages} {084709}
  (\bibinfo {year} {2007})}\BibitemShut {NoStop}%
\bibitem [{\citenamefont {Glauber}(1963){\natexlab{b}}}]{Glauber:1963_P-Rep}%
  \BibitemOpen
  \bibfield  {author} {\bibinfo {author} {\bibfnamefont {R.~J.}\ \bibnamefont
  {Glauber}},\ }\href@noop {} {\bibfield  {journal} {\bibinfo  {journal} {Phys.
  Rev.},\ }\textbf {\bibinfo {volume} {131}},\ \bibinfo {pages} {2766}
  (\bibinfo {year} {1963}{\natexlab{b}})}\BibitemShut {NoStop}%
\bibitem [{\citenamefont {Sudarshan}(1963)}]{Sudarshan_1963_P-Rep}%
  \BibitemOpen
  \bibfield  {author} {\bibinfo {author} {\bibfnamefont {E.~C.~G.}\
  \bibnamefont {Sudarshan}},\ }\href@noop {} {\bibfield  {journal} {\bibinfo
  {journal} {Phys. Rev. Lett.},\ }\textbf {\bibinfo {volume} {10}},\ \bibinfo
  {pages} {277} (\bibinfo {year} {1963})}\BibitemShut {NoStop}%
\bibitem [{\citenamefont {Bowden}(1972)}]{Bowden:1972}%
  \BibitemOpen
  \bibfield  {author} {\bibinfo {author} {\bibfnamefont {C.~M.}\ \bibnamefont
  {Bowden}},\ }\href@noop {} {\bibfield  {journal} {\bibinfo  {journal} {Int.
  J. Quantum Chem.},\ }\textbf {\bibinfo {volume} {6}},\ \bibinfo {pages} {133}
  (\bibinfo {year} {1972})}\BibitemShut {NoStop}%
\bibitem [{\citenamefont
  {Perelomov}(1972)}]{Perelomov_1972_Coherent_states_LieG}%
  \BibitemOpen
  \bibfield  {author} {\bibinfo {author} {\bibfnamefont {A.~M.}\ \bibnamefont
  {Perelomov}},\ }\href@noop {} {\bibfield  {journal} {\bibinfo  {journal}
  {Commun. Math. Phys.},\ }\textbf {\bibinfo {volume} {26}},\ \bibinfo {pages}
  {222} (\bibinfo {year} {1972})}\BibitemShut {NoStop}%
\bibitem [{\citenamefont {Gilmore}(1974)}]{Gilmore:1974}%
  \BibitemOpen
  \bibfield  {author} {\bibinfo {author} {\bibfnamefont {R.}~\bibnamefont
  {Gilmore}},\ }\href@noop {} {\bibfield  {journal} {\bibinfo  {journal} {Rev.
  Mex. Fis.},\ }\textbf {\bibinfo {volume} {23}},\ \bibinfo {pages} {143}
  (\bibinfo {year} {1974})}\BibitemShut {NoStop}%
\bibitem [{\citenamefont {Perelomov}(1977)}]{Perelomov_GCS_bosons}%
  \BibitemOpen
  \bibfield  {author} {\bibinfo {author} {\bibfnamefont {A.~M.}\ \bibnamefont
  {Perelomov}},\ }\href@noop {} {\bibfield  {journal} {\bibinfo  {journal}
  {Sov. Phys. Usp.},\ }\textbf {\bibinfo {volume} {20}},\ \bibinfo {pages}
  {703} (\bibinfo {year} {1977})}\BibitemShut {NoStop}%
\bibitem [{\citenamefont {Klauder}\ and\ \citenamefont
  {Skagerstam}(1985)}]{Klauder:1985}%
  \BibitemOpen
  \bibfield  {author} {\bibinfo {author} {\bibfnamefont {J.~R.}\ \bibnamefont
  {Klauder}}\ and\ \bibinfo {author} {\bibfnamefont {B.-S.}\ \bibnamefont
  {Skagerstam}},\ }\href@noop {} {\emph {\bibinfo {title} {{Coherent states:
  Applications in physics and mathematical physics}}}}\ (\bibinfo  {publisher}
  {World Scientific, Singapore},\ \bibinfo {year} {1985})\BibitemShut {NoStop}%
\bibitem [{\citenamefont {Cahill}\ and\ \citenamefont
  {Glauber}(1999)}]{Cahill:1999}%
  \BibitemOpen
  \bibfield  {author} {\bibinfo {author} {\bibfnamefont {K.~E.}\ \bibnamefont
  {Cahill}}\ and\ \bibinfo {author} {\bibfnamefont {R.~J.}\ \bibnamefont
  {Glauber}},\ }\href@noop {} {\bibfield  {journal} {\bibinfo  {journal} {Phys.
  Rev. A},\ }\textbf {\bibinfo {volume} {59}},\ \bibinfo {pages} {1538}
  (\bibinfo {year} {1999})}\BibitemShut {NoStop}%
\bibitem [{\citenamefont {{\"O}gren}\ \emph {et~al.}(2010)\citenamefont
  {{\"O}gren}, \citenamefont {Kheruntsyan},\ and\ \citenamefont
  {Corney}}]{Ogren:2010_Qdynamics_fermions_MolDiss}%
  \BibitemOpen
  \bibfield  {author} {\bibinfo {author} {\bibfnamefont {M.}~\bibnamefont
  {{\"O}gren}}, \bibinfo {author} {\bibfnamefont {K.}~\bibnamefont
  {Kheruntsyan}}, \ and\ \bibinfo {author} {\bibfnamefont {J.~F.}\ \bibnamefont
  {Corney}},\ }\href@noop {} {\bibfield  {journal} {\bibinfo  {journal}
  {EuroPhys. Lett.},\ }\textbf {\bibinfo {volume} {92}},\ \bibinfo {pages}
  {36003} (\bibinfo {year} {2010})}\BibitemShut {NoStop}%
\bibitem [{\citenamefont {{\"O}gren}\ \emph {et~al.}(2011)\citenamefont
  {{\"O}gren}, \citenamefont {Kheruntsyan},\ and\ \citenamefont
  {Corney}}]{Ogren2011_fermiondynamics}%
  \BibitemOpen
  \bibfield  {author} {\bibinfo {author} {\bibfnamefont {M.}~\bibnamefont
  {{\"O}gren}}, \bibinfo {author} {\bibfnamefont {K.}~\bibnamefont
  {Kheruntsyan}}, \ and\ \bibinfo {author} {\bibfnamefont {J.~F.}\ \bibnamefont
  {Corney}},\ }\href@noop {} {\bibfield  {journal} {\bibinfo  {journal}
  {Comput. Phys. Commun.},\ }\textbf {\bibinfo {volume} {182}},\ \bibinfo
  {pages} {1999} (\bibinfo {year} {2011})}\BibitemShut {NoStop}%
\bibitem [{\citenamefont {Rosales-Z\'arate}\ and\ \citenamefont
  {Drummond}(2011)}]{PRA_Entropy_paper_2011}%
  \BibitemOpen
  \bibfield  {author} {\bibinfo {author} {\bibfnamefont {L.~E.~C.}\
  \bibnamefont {Rosales-Z\'arate}}\ and\ \bibinfo {author} {\bibfnamefont
  {P.~D.}\ \bibnamefont {Drummond}},\ }\href@noop {} {\bibfield  {journal}
  {\bibinfo  {journal} {Phys. Rev. A},\ }\textbf {\bibinfo {volume} {84}},\
  \bibinfo {pages} {042114} (\bibinfo {year} {2011})}\BibitemShut {NoStop}%
\bibitem [{\citenamefont {Ma}\ and\ \citenamefont {Zhang}(1995)}]{Ma:1995}%
  \BibitemOpen
  \bibfield  {author} {\bibinfo {author} {\bibfnamefont {L.}~\bibnamefont
  {Ma}}\ and\ \bibinfo {author} {\bibfnamefont {Y.-d.}\ \bibnamefont {Zhang}},\
  }\Doi {10.1007/BF02726156} {\bibfield  {journal} {\bibinfo  {journal} {Il
  Nuovo Cimento B Series 11},\ }\textbf {\bibinfo {volume} {110}},\ \bibinfo
  {pages} {1103} (\bibinfo {year} {1995})}\BibitemShut {NoStop}%
\bibitem [{\citenamefont {Dyson}(1962)}]{Dyson:1962_Threefold}%
  \BibitemOpen
  \bibfield  {author} {\bibinfo {author} {\bibfnamefont {F.~J.}\ \bibnamefont
  {Dyson}},\ }\href@noop {} {\bibfield  {journal} {\bibinfo  {journal} {J.
  Math. Phys.},\ }\textbf {\bibinfo {volume} {3}},\ \bibinfo {pages} {1199}
  (\bibinfo {year} {1962})}\BibitemShut {NoStop}%
\bibitem [{\citenamefont {Wybourne}(1974)}]{Wybourne_Group_theory_book}%
  \BibitemOpen
  \bibfield  {author} {\bibinfo {author} {\bibfnamefont {B.~G.}\ \bibnamefont
  {Wybourne}},\ }\href@noop {} {\emph {\bibinfo {title} {Classical groups for
  physicists}}}\ (\bibinfo  {publisher} {Wiley, New York},\ \bibinfo {year}
  {1974})\BibitemShut {NoStop}%
\bibitem [{\citenamefont {Balian}\ and\ \citenamefont
  {Brezin}(1969)}]{Balian:1969}%
  \BibitemOpen
  \bibfield  {author} {\bibinfo {author} {\bibfnamefont {R.}~\bibnamefont
  {Balian}}\ and\ \bibinfo {author} {\bibfnamefont {E.}~\bibnamefont
  {Brezin}},\ }\href@noop {} {\bibfield  {journal} {\bibinfo  {journal} {Il
  Nuovo Cimento B},\ }\textbf {\bibinfo {volume} {64}},\ \bibinfo {pages} {37}
  (\bibinfo {year} {1969})}\BibitemShut {NoStop}%
\bibitem [{\citenamefont {Cartan}(1935)}]{Cartan:1935}%
  \BibitemOpen
  \bibfield  {author} {\bibinfo {author} {\bibfnamefont {E.}~\bibnamefont
  {Cartan}},\ }\href@noop {} {\bibfield  {journal} {\bibinfo  {journal}
  {Abhandlungen aus dem Mathematischen Seminar der Universitat Hamburg},\
  }\textbf {\bibinfo {volume} {11}},\ \bibinfo {pages} {116} (\bibinfo {year}
  {1935})}\BibitemShut {NoStop}%
\bibitem [{\citenamefont {Hua}(1963)}]{Hua:1963}%
  \BibitemOpen
  \bibfield  {author} {\bibinfo {author} {\bibfnamefont {L.-K.}\ \bibnamefont
  {Hua}},\ }\href@noop {} {\emph {\bibinfo {title} {Harmonic analysis of
  functions of several complex variables in the classical domains}}}\ (\bibinfo
   {publisher} {American Mathematical Society, Providence, Rhode Island},\
  \bibinfo {year} {1963})\BibitemShut {NoStop}%
\bibitem [{\citenamefont {Helgason}(2001)}]{Helgason:2001}%
  \BibitemOpen
  \bibfield  {author} {\bibinfo {author} {\bibfnamefont {S.}~\bibnamefont
  {Helgason}},\ }\href@noop {} {\emph {\bibinfo {title} {{ Differential
  geometry, Lie groups, and symmetric spaces}}}}\ (\bibinfo  {publisher}
  {American Mathematical Society},\ \bibinfo {address} {Providence, R.I.},\
  \bibinfo {year} {2001})\BibitemShut {NoStop}%
\bibitem [{\citenamefont {Hurwitz}(1897)}]{Hurwitz:1897}%
  \BibitemOpen
  \bibfield  {author} {\bibinfo {author} {\bibfnamefont {A.}~\bibnamefont
  {Hurwitz}},\ }\href {http://eudml.org/doc/58378} {\bibfield  {journal}
  {\bibinfo  {journal} {Nachr. Ges. Wiss. G{\"{o}}ttingen},\ }\textbf {\bibinfo
  {volume} {1897}},\ \bibinfo {pages} {71} (\bibinfo {year}
  {1897})}\BibitemShut {NoStop}%
\bibitem [{\citenamefont {Forrester}()}]{Forrester}%
  \BibitemOpen
  \bibfield  {author} {\bibinfo {author} {\bibfnamefont {P.~J.}\ \bibnamefont
  {Forrester}},\ }\href@noop {} {}\bibinfo {howpublished} {private
  communication}\BibitemShut {NoStop}%
\bibitem [{\citenamefont {Fyodorov}(2005)}]{FyodorovMetric}%
  \BibitemOpen
  \bibfield  {author} {\bibinfo {author} {\bibfnamefont {Y.~V.}\ \bibnamefont
  {Fyodorov}},\ }in\ \href@noop {} {\emph {\bibinfo {booktitle} {Recent
  perspectives in random matrix theory and number theory}}},\ \bibinfo {editor}
  {edited by\ \bibinfo {editor} {\bibfnamefont {F.}~\bibnamefont {Mezzadri}}\
  and\ \bibinfo {editor} {\bibfnamefont {N.~C.}\ \bibnamefont {Snaith}}}\
  (\bibinfo  {publisher} {Cambridge University Press},\ \bibinfo {address}
  {Cambridge},\ \bibinfo {year} {2005})\BibitemShut {NoStop}%
\bibitem [{\citenamefont {Forrester}(2010)}]{Forrester:2010}%
  \BibitemOpen
  \bibfield  {author} {\bibinfo {author} {\bibfnamefont {P.~J.}\ \bibnamefont
  {Forrester}},\ }\href@noop {} {\emph {\bibinfo {title} {{Log-gases and random
  matrices }}}}\ (\bibinfo  {publisher} {Princeton University Press},\ \bibinfo
  {address} {Princeton},\ \bibinfo {year} {2010})\BibitemShut {NoStop}%
\bibitem [{\citenamefont {Mehta}(2004)}]{Mehta:2004}%
  \BibitemOpen
  \bibfield  {author} {\bibinfo {author} {\bibfnamefont {M.~L.}\ \bibnamefont
  {Mehta}},\ }\href@noop {} {\emph {\bibinfo {title} {Random matrices}}},\
  \bibinfo {edition} {3rd}\ ed.\ (\bibinfo  {publisher} {Academic Press,
  Boston},\ \bibinfo {year} {2004})\BibitemShut {NoStop}%
\bibitem [{\citenamefont {Fan}(2007)}]{Fan:2007}%
  \BibitemOpen
  \bibfield  {author} {\bibinfo {author} {\bibfnamefont {H.-Y.}\ \bibnamefont
  {Fan}},\ }\href@noop {} {\bibfield  {journal} {\bibinfo  {journal} {Ann.
  Phys.},\ }\textbf {\bibinfo {volume} {322}},\ \bibinfo {pages} {886}
  (\bibinfo {year} {2007})}\BibitemShut {NoStop}%
\bibitem [{\citenamefont {Ohnuki}\ and\ \citenamefont
  {Kashiwa}(1978)}]{Ohnuki_1978_CSFermiOp_PathInt}%
  \BibitemOpen
  \bibfield  {author} {\bibinfo {author} {\bibfnamefont {Y.}~\bibnamefont
  {Ohnuki}}\ and\ \bibinfo {author} {\bibfnamefont {T.}~\bibnamefont
  {Kashiwa}},\ }\href@noop {} {\bibfield  {journal} {\bibinfo  {journal} {Prog.
  Theor. Phys.},\ }\textbf {\bibinfo {volume} {60}},\ \bibinfo {pages} {548}
  (\bibinfo {year} {1978})}\BibitemShut {NoStop}%
\end{thebibliography}

\end{document}